\documentclass[pre,reprint,twocolumn,tightenlines,nopacs,showkeys,superscriptaddress,groupedaddress]{revtex4} 
\usepackage{graphicx, bm} 
\usepackage{dcolumn} 
\usepackage{bm} 
\usepackage{amssymb} 
\usepackage{comment} 
\usepackage[caption=false]{subfig}
\usepackage{amsmath}
\usepackage{float} 
\usepackage{natbib} 
\usepackage{tabularx}
\usepackage{dsfont} 
\usepackage{cancel}
\usepackage{color}

\newcommand*\xbar[1]{%
  \hbox{%
    \vbox{%
      \hrule height 0.5pt 
      \kern0.5ex
      \hbox{%
        \kern-0.1em
        \ensuremath{#1}%
        \kern-0.1em
      }%
    }%
  }%
} 

\begin{document}

\title{What the odor is not: Estimation by elimination}

 \author{Vijay Singh} \affiliation{Department of Physics, North Carolina A\&T State University, Greensboro, NC, 27410, USA} \affiliation{Department of Physics, \& Computational Neuroscience Initiative, University of Pennsylvania, Philadelphia, PA 19104, USA}
\email{vsingh@ncat.edu}
\author{Martin Tchernookov} \affiliation{Department of Physics, University of Wisconsin, Whitewater, WI, 53190, USA}
\author{Vijay Balasubramanian} \affiliation{Department of Physics, \& Computational Neuroscience Initiative, University of Pennsylvania, Philadelphia, PA 19104, USA}
\email{vijay@sas.upenn.edu}

\begin{abstract}
Olfactory systems use a small number of broadly sensitive receptors to combinatorially encode a vast number of odors.  We propose a method of decoding such distributed representations by exploiting a statistical fact: receptors that do not respond to an odor carry more information than receptors that do because they signal the absence of all odorants that bind to them. Thus, it is easier to identify what the odor {\em is not}, rather than what the odor is. For  realistic numbers of receptors, response functions, and odor complexity, this method of elimination turns an underconstrained decoding problem into a solvable one, allowing accurate determination of odorants in a mixture and their concentrations. We construct a neural network realization of our algorithm based on the structure of the olfactory pathway.
\end{abstract}

\maketitle

\section{Introduction} 
The olfactory system enables animals to sense, perceive, and respond to mixtures of volatile molecules  carrying messages about the world.   There are perhaps $10^4$ or more monomolecular odorants \cite{dunkel2008superscent,touhara2009sensing,yu2015drawing}, far more than the number of receptor types in animals ($\sim50$ in fly, $\sim300$ in human, $\sim1000$ in rat, mouse and dog \cite{vosshall2000olfactory, zozulya2001human, zhang2002olfactory, quignon2005dog}).  The problem of representing high-dimensional chemical space in  a low-dimensional response space may be solved by the presence of many receptors that bind to numerous odorants \cite{audouze2014identification, saito2009odor, mainland2015human, araneda2000molecular, malnic1999combinatorial,su2009olfactory, mainland2014molecule}, leading to a distributed, compressed, and combinatorial representation \cite{malnic1999combinatorial,hopfield1999odor,bazhenov2009olfactory,stettler2009representations,zhang2016robust,qin2019optimal,kadakia2019front, assisi2020optimality} processed by  activity in networks of  neurons \cite{laurent2001odor, kay2006information}.  Some mechanisms for such distributed representation  propose that each odorant activates specific subsets of neurons \cite{koulakov2007olfactory, stevens2015fly,zwicker2016receptor,krishnamurthy2017disorder}.
Other models propose that the olfactory network assigns similar activity patterns to similar odors \cite{dasgupta2017neural,dasgupta2018neural} and classifies odors as activity clusters in an online and supervised manner \cite{pehlevan2017clustering}.
Population  models  suggest that odor identity and intensity could be represented  in dynamical response patterns  \cite{stopfer2003intensity,sanda2016classification}, where different odors activate distinct  attractor network patterns in a winner-less competition \cite{rabinovich2000dynamical,laurent2001odor,pehlevan2017clustering}, or in  transient \cite{mazor2005transient} or oscillatory activity \cite{laurent1996dynamical}. Finally, population activity could be a low-dimensional projection of  odor space \cite{turner2008olfactory,raman2011mimicking} evolving  in space and time to decorrelate  odors  \cite{laurent2002olfactory} to   maximally separate sparse representations of similar odors \cite{assisi2020optimality}.

Here, we focus on a simplified inverse problem: odor composition estimation from time-averaged, combinatorial receptor responses.  
We thus omit  receptor and circuit dynamics important in many olfactory phenomena in animals  to concentrate   on odor sensing combinatorics (also see \cite{babadi2014sparseness,zhang2016robust,stevens2015fly,zwicker2016receptor, krishnamurthy2017disorder,dasgupta2017neural, dasgupta2018neural,Tesileanu255547}).
We propose that receptors that {\em do not} respond to an odor carry more information about it than receptors that {\it do}. 
This is because silent receptors signal that none of the odorants that could bind to them  are present.  Most absent odorants can be identified and eliminated with just a few such silent receptors. Thus, it is easier to identify what the mixture {\em is not}, rather than what the mixture is. For realistic parameters, this elimination turns odor composition estimation from an underdetermined to an overdetermined problem. Then,  the remaining odorants can be estimated from active receptor responses.

To be specific, we use  realistic competitive binding models of odor encoding by receptors \cite{rospars2008competitive, cruz2013neural, reddy2018antagonism, singh2019competitive},  and propose schemes to estimate odor composition from such responses. The schemes work over a  range of 
parameters, do not require  special constraints on receptor-odorant interactions, and work for systems with few receptors sensing odors of natural complexity.
We then develop a neural network inspired by the known structure of the olfactory system to decode odors from  receptor responses.
We provide performance bounds for these decoders on standard tasks such as detecting  presence or absence of odorants in mixtures \cite{rokni2014olfactory}, and discriminating  mixtures that differ in some components \cite{jinks1999limit, bushdid2014humans}.  
These algorithmic schemes are designed with prior knowledge of receptor responses to the complete space of relevant odorants, and hence do not apply as presented to biological olfaction. However, we also  construct a version of our neural network  without such prior information, albeit at the price of producing a sparse, distributed representation of odors which must be subsequently decoded by a trained classifier.

\section{Results}
\subsection{Identifying odorant presence}
Suppose we just seek to identify  presence or absence of  odor components and not  concentrations because 
the  task requires it, or when receptor noise is high.
In the latter case, stochastic binding dynamics makes  exact binding states hard to predict and receptor activation is determined by noise thresholds.
Either way, if the concentration is high and evokes above-threshold receptor activity, the odorant is considered present and the receptor is active.
Otherwise, the receptor is deemed inactive and the odorant is absent.
Key features of our scheme can be explained in this model.
Later, we consider realistic competitive binding (CB) models,= that include continuum odorant concentrations and receptor responses, and then network models which do and do not assume knowledge of the number of odorants and sensing matrix.

Consider mixture of $N_{\rm L}$ odorants represented by  binary vector $\boldsymbol{c} = (c_1,c_2,\hdots,c_{N_{\rm L}})$, where $c_i=1$ represents presence of the $i$'th odorant.    Suppose that  only $K$ odorants are present in the mixture on average.
These odorants bind to $N_{\rm R}$ receptors whose response is given by the vector $\boldsymbol{R} = (R_1,R_2,\hdots,R_{N_{\rm R}})$.
Receptor sensitivity to  odorants is given by a matrix $S$, which we assume  known. 
$S_{ij} = 1$ indicates that the odorant $j$ can bind to receptor $i$ and $S_{ij}=0$ means it can not.  
Suppose that the probability that an odorant binds to a receptor is $s$, i.e., $P(S_{ij} =1) = s$.
Then, on average, each odorant binds to $s N_{\rm R}$ receptors and each receptor to $s N_{\rm L}$ odorants.  

Here receptors respond ($R_i = 1$) to odors containing at least one odorant  binding to them.
Without such an odorant, the receptor is inactive ($R_i = 0$).
The receptor thus acts as an `OR' gate, approximating a biophysical 
model \cite{singh2019competitive, rospars2008competitive, cruz2013neural, reddy2018antagonism},
with a sigmoidal response function (see below)
in situations with high concentration odorants or a sharp threshold and steep response.

\begin{figure}[t]
\begin{center}
\includegraphics[width=\linewidth]{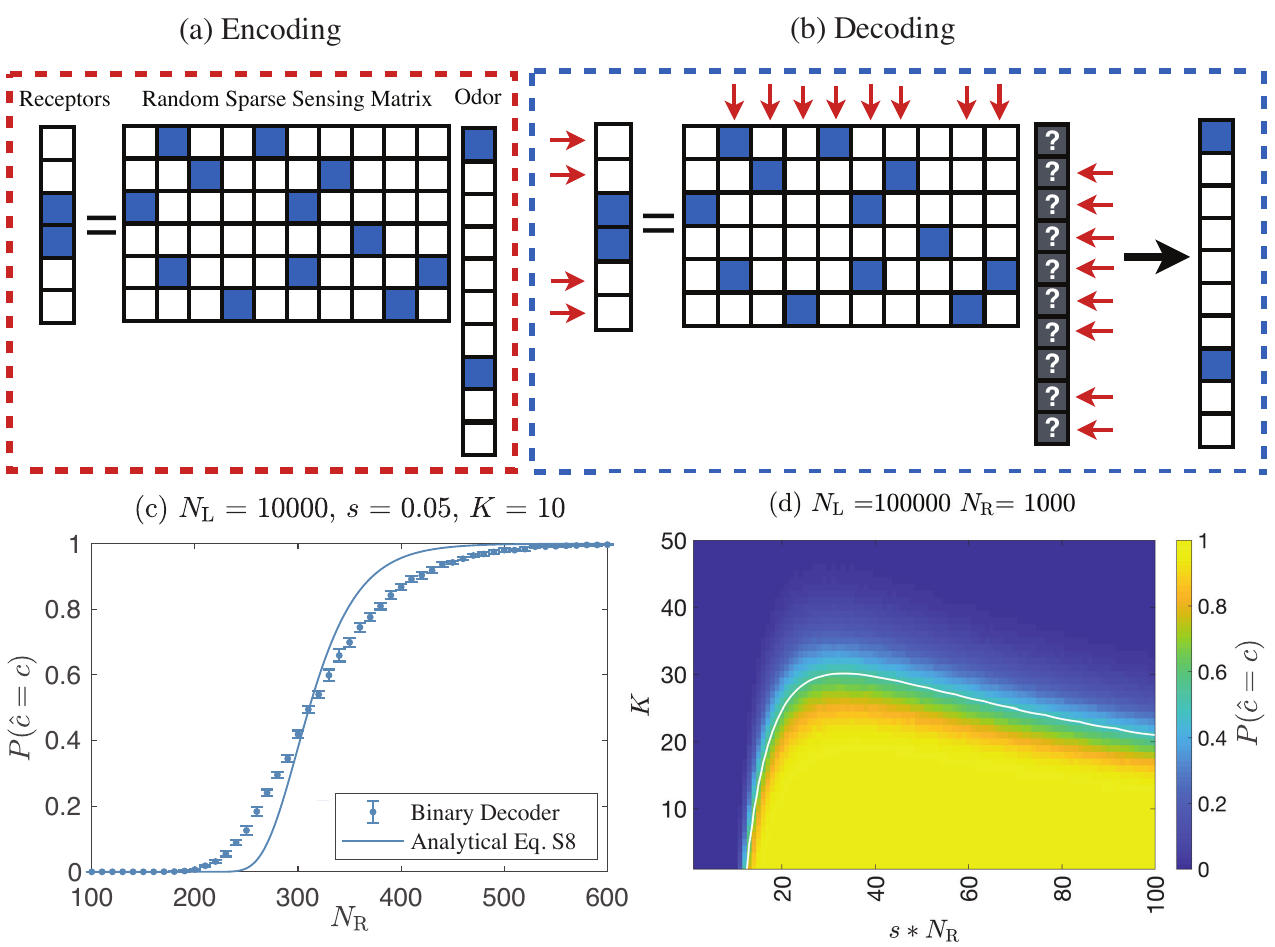} 
\end{center}
    \caption{{\bf Identifying odorant presence:} ({\bf a}) Encoding: The odorant mixture and receptor activity are effectively binary vectors. If an odorant  drives receptor activity  above threshold, the odorant is considered  present and the receptor is said to respond; otherwise the odorant is absent and the receptor is non-responsive. Present odorants and active receptors  are indicated as filled  elements of  corresponding vectors. Here, odorant 1 and 8 are present and receptors 3 and 4 respond.  Receptor response is obtained from the sensitivity matrix  (filled  elements indicate  odorant-receptor  binding). Here, receptor 1 binds to odorants 2 and 5; receptor 2 to odorants 3 and 7; etc. A receptor is active if the odor contains at least one odorant  binding to it; otherwise the receptor is inactive. ({\bf b}) Decoding: (Step 1) Absent odorants  are identified from inactive receptors.  (Step 2)  Remaining odorants are considered present. 
({\bf c}) Correct decoding probability ($P(\hat{{\boldsymbol{c}}} = {\boldsymbol{c}})$)  as a function of number of receptors ($N_{\rm R}$). Markers = simulations; smooth curve =   analytical result (Eq.~S8).   $P(\hat{{\boldsymbol{c}}} = {\boldsymbol{c}})$ measured as a fraction of correct decodings over 1000 trials  with random choices of odor mixture and sensitivity matrix (see SI \cite{suppMat}).  Mean and error bar ($\pm1$ standard deviation)  computed over 10 replicate simulations  (1000 trials each).
     ({\bf d}) $P(\hat{{\boldsymbol{c}}} = {\boldsymbol{c}})$  as a function of the average number of odorants in mixtures ($K$) and the average number of receptors responding to an odorant ($s*N_{\rm R}$) (number of receptors ($N_{\rm R}$) and odorants ($N_L$)  fixed).  $P(\hat{{\boldsymbol{c}}} = {\boldsymbol{c}})$ is plotted as a function of $s*N_{\rm R}$, since  $s$ and $N_{\rm R}$  appear in this combination (Eq.~\ref{eq:BooleanProb4} and Eq.~S8). $s*N_{\rm R}$  (average number of receptors binding to an odorant), determines whether the odorant is detectable. $P(\hat{{\boldsymbol{c}}} = {\boldsymbol{c}})$ is measured over 10000 trials  with random choices of odor mixture and sensitivity matrix. The  white curve is 
     estimated by setting the analytical expression in 
     Eq.~\ref{eq:BooleanProb4} to 0.5.
     }
\label{fig:binaryScheme}
  \end{figure}

Odors encoded in this way can be decoded (estimate $\boldsymbol{\hat{c}}$) in two  steps (Fig.~\ref{fig:binaryScheme}): ({\bf 1}) Identify inactive receptors and declare odorants that bind to these  receptors as absent, and ({\bf 2})  Declare the remaining odorants  present.
This  decoder identifies all odorants  in the mixture because, assuming  odorants bind to at least one receptor, all receptors that bind to an odorant that is present ($c_j=1)$  will respond.  Hence, its presence will be identified.

False positives are possible because receptors that bind to an absent odorant could have  non-zero response because of other odorants present in the mixture.
Thus, the missing odorant  will be declared present, giving a false positive.
The probability of such false positives is approximately
(see SI: Identifying odorant presence \cite{suppMat}):
\begin{equation}
P(\hat{c}_i=1| c_i =0) \approx e^{-s N_{\rm R} e^{-s K}}.
\label{eq:falsePositives}
\end{equation}
We can  derive an approximate probability of correct estimation assuming that each odorant is estimated independently of others 
(SI: Identifying odorant presence Eq.~S8 \cite{suppMat}):
\begin{equation}
P(\hat{{\boldsymbol{c}}} = {\boldsymbol{c}})  \approx \left(1 - {N_{\rm L}} e^{-s N_{\rm R} e^{-s K}}\right).
\label{eq:BooleanProb4}
\end{equation}
The second term  is the approximate  probability of false positives with $N_{\rm L}$ possible odorants in the environment.

For correct decoding the  false positive probability should be low;
so, the term in the exponent of Eq.~\ref{eq:falsePositives} should be large, i.e., $s N_{\rm R}$ should be large and $s K$ small.
This makes sense, as 
$s N_{\rm R}$ is the average number of receptors that an odorant binds. 
Thus, $s N_{\rm R}$ should be large so that
 many receptors can provide evidence for absence of the odorant by not responding.
Also, sufficiently many receptors must be inactive to eliminate all odorants that are absent.  For this to happen, the probability that any particular receptor responds to at least one of the K odorants in the mixture should be small.  This probability is  $\approx sK$ when the likelihood $s$ that a given odorant binds to a  receptor is small; so we require that $sK < 1$.

The conditions  $s N_{\rm R} > 1$ and $s K < 1$ are needed 
because an odorant's concentration cannot be estimated if it does not bind to any receptor, while converting an  under-determined problem into a well determined one requires sufficiently many inactive receptors.
Put otherwise, for fixed numbers of receptors and odorants, with fixed odor component complexity $K$, receptor sensitivity should be sufficiently high to ensure coverage of odorants, but small enough to avoid false positives.

These considerations can be combined with the observed sensitivity of olfactory receptors
($s\sim 5\%$ for mammals \cite{mainland2015human}).
For typical mammalian parameters ($\{N_{\rm L},  K , N_{\rm R}, s\} \sim \{10^4, 10, 500, 0.05\}$), 
the estimated false positive probability is low ($P(\hat{c}_i=1| c_i =0)\sim10^{-7}$; Eq.~\ref{eq:falsePositives}) and the  correct estimate probability is high ($P(\hat{{\boldsymbol{c}}} = {\boldsymbol{c}}) \sim 0.998$; Eq.~\ref{eq:BooleanProb4}).  These are  upper bounds because we considered binary, noiseless signals, while computation in the brain is degraded by noise in sensory and decision  circuitry and by circuit constraints, as discussed below.
However, our result here shows that in principle, and ignoring noise,  odor composition is fully recoverable from the sort of combinatorial codes implemented in the nose.

We  estimated our scheme's accuracy using sensitivity matrices with elements taken non-zero with probability $s$, i.e. ($P(S_{ij}>0)=s$).
Fig.~\ref{fig:binaryScheme}c shows the correct estimate probability ($P(\hat{{\boldsymbol{c}}}={\boldsymbol{c}})$)
as a function of the receptor number  $(N_{\rm R})$ for odors containing on average $K=10$ odorants drawn from $N_L = 10000$ possibilities.
When the number of receptors  is low,  the  correct decoding probability is zero.  As the number of receptors increases,  we transition to a region where recovery is good. The transition is sharp and occurs when the number of receptors is much smaller than the number of  odorants. Our  scheme performs well over a range of odor complexities ($K$) and numbers of responsive receptors ($s*N_{\rm R}$), 
(Fig.~\ref{fig:binaryScheme}d). Our  expression for the  correct decoding probability (SI  Eq.~S8 \cite{suppMat}) describes  the numerical results well, and gives a good estimate of the transition  between poor and good decoding as a function of odor complexity and  receptor sensitivity (solid lines in Fig.~\ref{fig:binaryScheme}c,d;  SI Fig.~S1 \cite{suppMat}).
We will show that including noise  reduces decoding accuracy, matching experiments, but the prediction of a sharp threshold between poor and good performance remains.

\begin{figure*}
\begin{center} 
\includegraphics[width=0.9\linewidth]{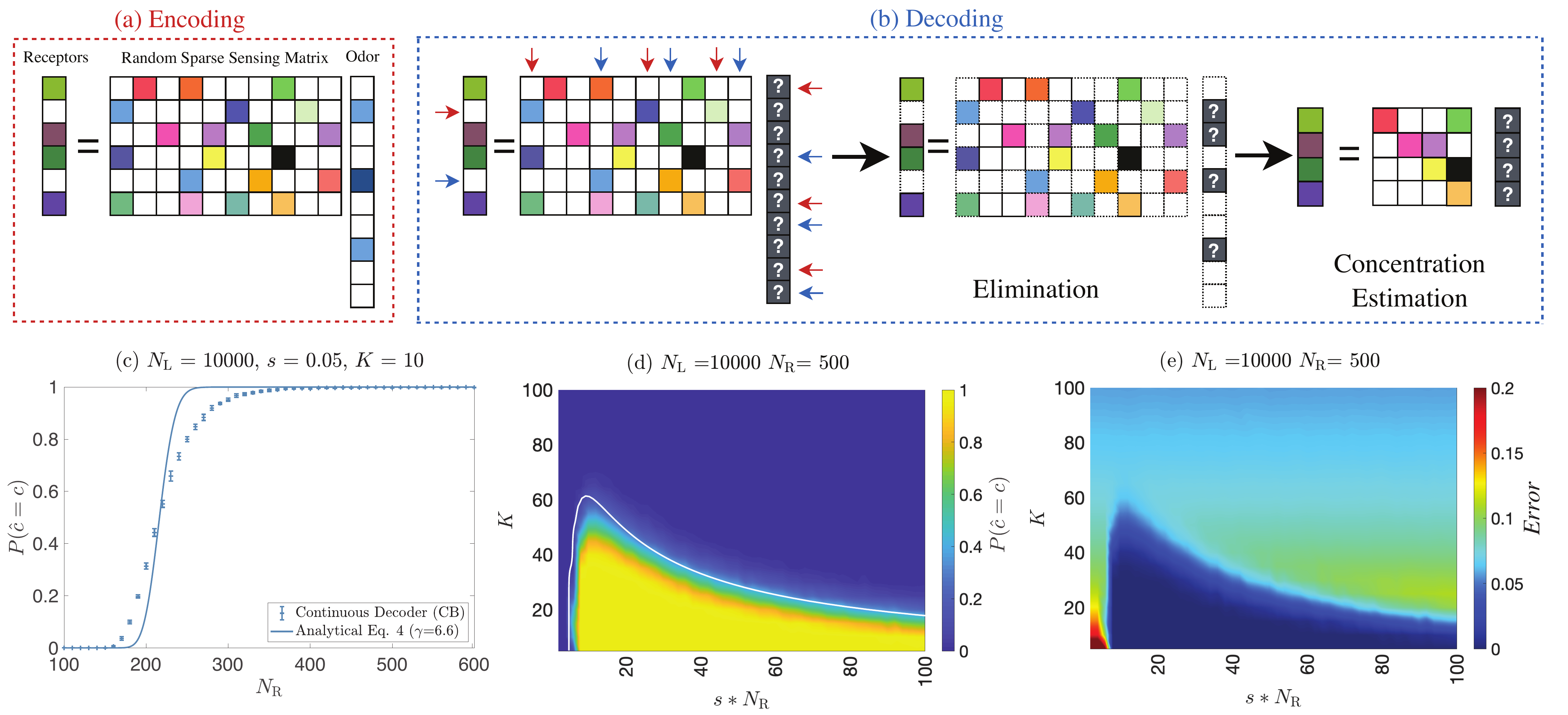} 
    \caption{{\bf Identifying odorant concentrations:}
({\bf a}) Encoding: Odor mixtures are $N_{\rm L}$ component vectors with $K$ non-zero entries  represented by  filled  elements (color saturation = concentration). 
Elements of the sensing matrix (rows = receptors, columns = odorants) indicate  strength of receptor-odor interactions (color = receptor binding affinity to an odorant; white = zero affinity). 
Receptor activity = colors, white =  inactive.  
({\bf b}) Decoding (Step 1 Elimination): Inactive receptors are used to eliminate absent odorants, reducing an under-determined problem to a well-defined one. (Step 2 Estimation):  Concentrations of  remaining odorants are estimated from responses of active receptors. 
({\bf c}) Same as Fig.~\ref{fig:binaryScheme}(c), now for the continuous decoder.  Mean and  error bar ($\pm1$ s.d.)  over 10 replicate simulations, each with 1000 trials.  The  parameter $\gamma$ in the  (Eq.~\ref{eq:theoreticalLimitMain}) was chosen to minimize MSE between the numerical probability and the formula. 
({\bf d}) $P(\hat{{\boldsymbol{c}}} = {\boldsymbol{c}})$  as a function of number of odorants  ($K$) and $s*N_{\rm R}$ at fixed $N_{\rm R}$. $P(\hat{{\boldsymbol{c}}} = {\boldsymbol{c}})$  calculated  over 1000 trials, each with random choices of odor mixture and sensitivity matrix. The white curve is the  boundary of the good decoding region, estimated by setting Eq.~\ref{eq:theoreticalLimitMain} to $0.5$ ($\gamma = 3$).
({\bf e}) Concentration estimate error (Euclidean distance between actual and estimated  values divided by  number of odorants; $\left( ||\hat{\boldsymbol{c}} - {\boldsymbol{c}}||_2 / K \right)$),   as a function of number of odorants  ($K$) and $s*N_{\rm R}$ at fixed $N_{\rm R}$.   Error is low even when recovery is imperfect.  Other error measures give similar results  (SI Fig.~S3: Other measures of estimation error \cite{suppMat}).
    }
\label{fig:continuousScheme}
  \end{center} 
  \end{figure*}

\subsection{Identifying odorant concentrations}
When noise is low, or integration times are long, fine gradations of  odorant concentrations ($c_i$), receptor sensitivities ($S_{ij}$), and receptor responses ($R_i$) can be discriminated.
The response  is then well-described as a Hill function \cite{singh2019competitive, reddy2018antagonism, rospars2008competitive, cruz2013neural} because odorant molecules compete to occupy receptor binding sites \cite{singh2019competitive}:
\begin{equation}
R_i = \frac{\sum_{j=1}^{N_{\rm L}} 
S_{ij}c_j
}{\left(1+d*\sum_{j=1}^{N_{\rm L}} 
S_{ij}c_j
\right)},
\label{eq:compBind}
\end{equation}
where $c_i$ are odorant concentrations and 
$d$ parameterizes  affinity for the receptor. The response is binary when $d$ is large ($R=0$ or $1/d$), and linear \cite{gupta2015olfactory,martelli2013intensity, ferreira2012revisiting,Tesileanu255547} when $d\rightarrow0$.  Synergy, suppression, antagonism, and inhibition, which may be widespread \cite{pfister2019odorant}, can be included  in this model  \cite{singh2019competitive}. Below we will first consider a decoder which has explicit knowledge of this receptor response model, and later present neural network models which do not assume knowledge of the number of odorants and response model.

Our decoder  can now be modified to estimate both which odorants are present and their concentrations.   We start with an under-determined  problem because the number of  odorants exceeds the number of receptors ($N_L > N_R$).   First, we eliminate odorants  binding to receptors with below-threshold responses. Thus, an odorant is considered functionally absent if its concentration is low enough that some receptors specific to it respond below threshold. This leaves $\tilde{N}_R$ active receptors responding to $\tilde{N}_L$ candidate odorants.   If  $\tilde{N}_{\rm L} \leq \tilde{N}_{\rm R}$ the problem is now over-determined and can be solved (Fig.~\ref{fig:continuousScheme}), even if some absent odorants have not been eliminated.   Specifically, we  invert the response functions (Eq.~\ref{eq:compBind})  relating the $\tilde{N}_L$ odorant concentrations to the  $\tilde{N}_R$  responses to get the unknown concentrations.

Our decoder will  eliminate none of the $K$ present odorants  because all 
evoke responses.  To estimate  false positives, 
let $s$ be the probability that a  receptor responds to a given odorant ($P(S_{ij}>0)=s$).  Then, the  number of active receptors is  $\tilde{N}_R \sim  s K N_{\rm R}$ while the number of inactive receptors is about $ (1 - sK) N_R$.  The first inactive receptor eliminates  roughly a fraction $s$ of the remaining 
$N_L - K$ 
odorants;  the second
removes   another fraction $s$ of the remaining $(1-s) (N_L - K)$ odorants.  Summing over these eliminations for all $(1 - sK) N_R$ inactive receptors leaves
$\tilde{N}_L \sim  K + (N_L - K)(1-s)^{N_R(1 - s K) - 1}$ odorants.
Typical  parameters $\{N_{\rm L},  K , N_{\rm R}, s\} = \{10^4, 10, 500, 0.05\}$ give $\tilde{N}_L \sim K = 10$ which is less than $\tilde{N}_R \sim  s K N_{\rm R} = 250$; so,  in the  relevant regime our  algorithm  leads to an over-determined and hence solvable identification problem.

We can find an approximate analytical expression for the probability of correct estimation (SI: Identifying odorant concentrations \cite{suppMat}) by assuming that the typical number of receptors  responding to a mixture exceeds the average odor complexity $(\tilde{N}_{\rm R} > K)$, while, at the same time, enough receptors are inactive to eliminate absent odorants. This requires $s(N_{\rm R}-\tilde{N}_{\rm R})>\gamma$, where $\gamma >1$ is a parameter depending on the  response model (SI: Identifying odorant concentrations \cite{suppMat}).
Then,
\begin{align}
P(\hat{{\boldsymbol{c}}} = {\boldsymbol{c}})  \sim P(\tilde{N}_{\rm R} > K)*P(N_{\rm R}-\tilde{N}_{\rm R}> (\gamma/s))
 \nonumber \\ = \left[ 1 - \Phi \left( \frac{K - \tilde{N}_{\rm R}}{\sqrt{\tilde{N}_{\rm R}}}\right)\right] 
\Phi \left( \frac{N_{\rm R} - \tilde{N}_{\rm R} - \frac{\gamma}{s}}{\sqrt{\tilde{N}_{\rm R}}}\right) \, .
\label{eq:theoreticalLimitMain}
\end{align}
$\Phi$ is the normal cumulative distribution function.

To  estimate the probability of correct decoding ($P(\hat{{\boldsymbol{c}}}={\boldsymbol{c}})$), we generated sparse odor vectors with  $K$ odorants on average. We drew concentrations from a uniform distribution on the interval [0, 1). Elements of the sensitivity matrix were chosen  non-zero with probability $s$;
non-zero values were chosen from a log uniform distribution (SI: Numerical Simulations; similar results with other distributions in SI Fig.~S2 \cite{suppMat}). With few receptors, the correct decoding probability vanishes. But the probability transitions sharply to finite values at a threshold $N_{\rm R}$ much smaller than the number of odorants 
(Fig.~\ref{fig:continuousScheme}c). Odor compositions are recovered well for a  range of parameters  (Fig.~\ref{fig:continuousScheme}d,e), if receptors are sufficiently sensitive  $s*N_{\rm R}>6$. Odors with the highest complexity are decoded when $s*N_{\rm R} \sim 7-15$. We quantified the error in odor estimates and found that even when decoding is not perfect there is a large parameter space where the error is small (Fig.~\ref{fig:continuousScheme}e). These results have weak dependence on the number of odorants (SI Fig.~S4 \cite{suppMat}).

With  $\sim 300$ receptors like human, our model predicts that odors with most components can be decoded with $s\sim 3-5\%$, while with $\sim 50$ receptors like {\it Drosophila} we need greater responsivity $s \gtrsim 13\%$ for best performance. Interestingly, human receptors have $s \sim 4\%$ \cite{mainland2015human} while in the fly $s \sim 14\%$ \cite{munch2016door}.

Our decoder can be modified to incorporate other biophysical interactions between odorants and receptors. If non-competitive interactions are known for a receptor-odor pair, the corresponding response model can be used instead of the CB model (Eq.~\ref{eq:compBind}). Odorant suppression of receptor responses can be included in the algorithm.  If responses are attenuated but not completely silenced, the algorithm goes through as before, albeit with a response model that includes the attenuation. Even if the suppressive interactions are strong, so long as such odor-receptor interactions are sparse, as experiments suggest \cite{saito2009odor, mainland2015human}, the algorithm can be modified to include receptor silencing by ignoring odorants and receptors  with strong suppressive interactions  in the elimination step, and  including them in the estimation step.  We leave a detailed investigation  to future work.

\begin{figure}
\begin{center} 
\includegraphics[width=\linewidth]{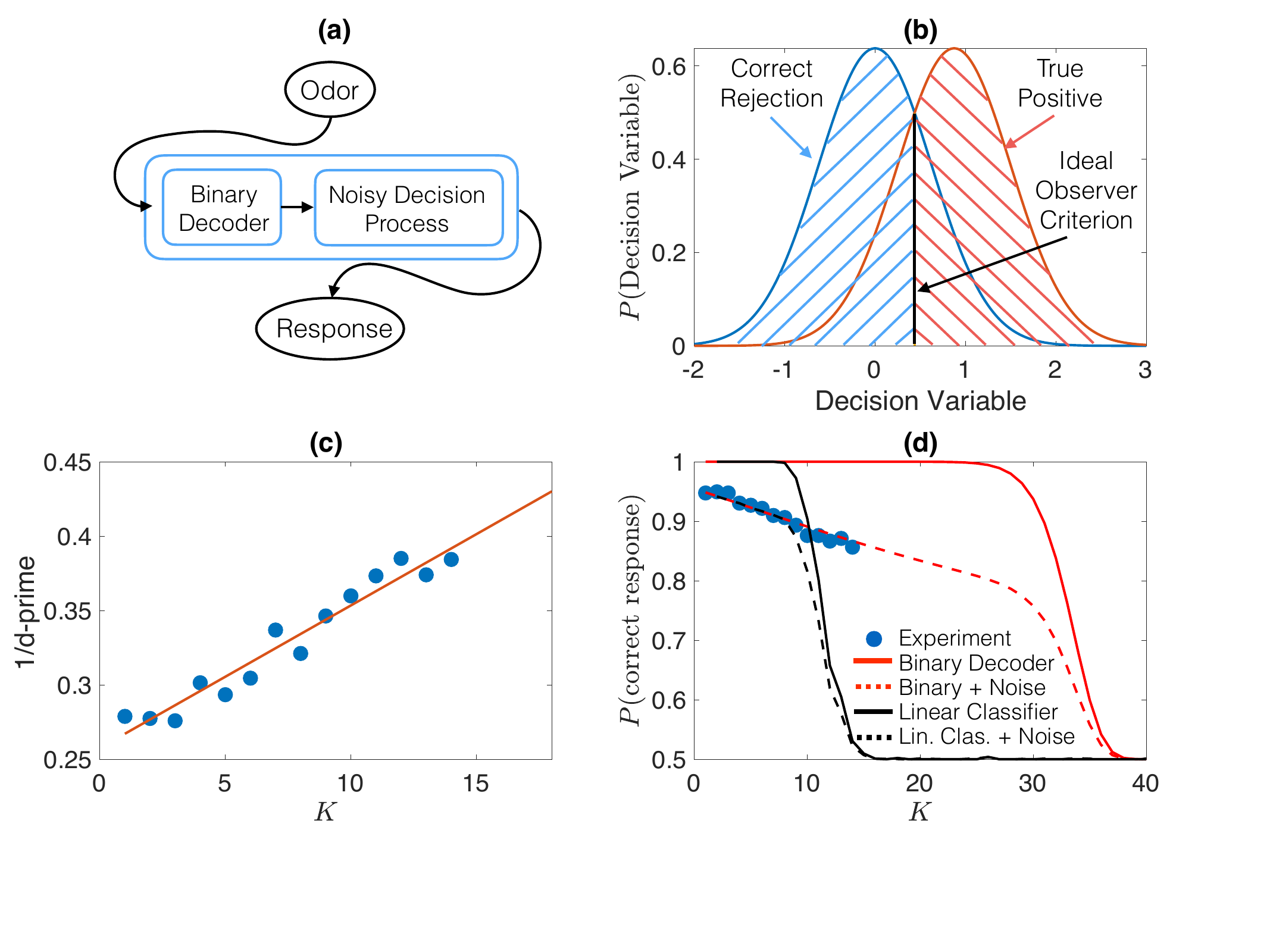} 
    \caption{ {\bf Effect of decision noise:} 
(a) Schematic of a two step noisy decoder. 
(b) We assume normally distributed decision variables with mean $0$ when the target is absent,  a mean equal to the probability of correct detection for the odorant presence decoder, $P(\hat{c}=c)$,  when the target is present, and identical standard deviations in both conditions. The ideal observer detection threshold is indicated. Probability of correct response equals  probability of correct rejection plus  probability of correct detection.  (c)  1/d-prime estimated from an olfactory cocktail-party task of detecting odorants in $K$-component mixtures  \cite{rokni2014olfactory} 
(details in text) follows a linear trend with $K$.   (d) Probability of correct detection of presence/absence of an odorant in a $K$-component mixture. Blue markers =  fraction of correct responses  (true positive + correct rejection).  Continuous lines = noiseless prediction for our decoder (red) vs. linear classifier (black).  Dashed lines = prediction including noise determined from   (c) for our decoder (red) vs. linear classifier (black).  Parameters:  number of odorants $N_{\rm L} = 10^4$, number of receptors $N_{\rm R}=1000$,  response  sensitivity $s=0.05$ \cite{mainland2015human}.
}
\label{fig:Test}
  \end{center} 
  \end{figure}

\subsection{Noise and decision making}

To study how noise degrades performance relative to an ideal decoder in our setting, we considered a ``cocktail-party problem'' where an agent seeks to identify presence or absence of a component in an odor mixture.  We then modeled noisy decision-making as a two-step process: (1) Internally representing  the mixture using the decoder of odor presence described above, and (2) Using noisy higher level processes to decide on  presence or absence of the target odorant based on the output of the estimation step (Fig.~\ref{fig:Test}a).

We modeled the noisy decision variables derived  from activity in a decision circuit
\cite{parker1998sense, gold2007neural}, by requiring that the baseline-subtracted decision variable should be $0$ for absent targets; for targets that are present, the variable should be proportional to the probability $P(\hat{c}=c) = p$ of correct detection. In both cases the decision variable is  distributed around the desired value with a standard deviation determined by  noise. An ideal observer  asks whether the target odorant is more likely to be present or absent, given the observed value of the decision variable and its distribution in the two cases (Fig.~\ref{fig:Test}b). 

Next, to derive a realistic noise model, we considered experiments where mice  were trained to report a target odorant in  mixtures of up to 14 of
16 odorants of identical concentration, any of which could be present or absent \cite{rokni2014olfactory}.
Mice reported presence/absence of all targets with high accuracy ($>80\%$), suggesting that they could learn to identify all mixture components. Decision noise  can be directly estimated from this data. Briefly,  for Gaussian decision variables (Fig.~\ref{fig:Test}b), we can estimate the standard deviation from the hit rate (fraction  of correct detections) and false alarm rate 
(fraction of incorrect detections). Signal detection theory \cite{green1966signal} relates signal to noise ratio (SNR; also called d-prime) of this go/no-go task to the hit/false alarm rates as: SNR = d-prime = z(hit) - z(false alarm), where z is the z-score.   This analysis gave the SNR for mice as a function of  the mixture complexity $K$ (\cite{rokni2014olfactory}; Fig.~\ref{fig:Test}c).  We estimated SNR at other values of $K$ by extrapolating the experimental relationship (Fig.~\ref{fig:Test}c, red line).   For a Gaussian decision variable with the same standard deviation $\sigma$ in both conditions, and a difference in means of $\mu$, standard theory  \cite{green1966signal} gives ${\rm SNR} = \mu /\sigma$. We took the noise standard deviation in our model to be  $\sigma$ (estimated from the data for each $K$) times a constant chosen to minimize the mean squared difference between theory and experiment.

Since we derived our noise model from data in mice, we constructed a decoder with $N_R=1000$ receptors and response sensitivity $s=0.05$ \cite{mainland2015human}.  Without noise ($\sigma = 0$), our  decoder predicts essentially perfect performance for identification of missing odorants in odors with up to $\sim27$ components, and a sharp fall-off thereafter (Fig.~\ref{fig:Test}d, red line).  Adding noisy decisions  (Fig.~\ref{fig:Test}a,b) leads to the dashed red line in Fig.~\ref{fig:Test}d.  Interestingly, there is a good match to the mouse behavioral data in  \cite{rokni2014olfactory} (RMSE between observed and predicted probability of correct estimate = 0.0075).  Based on our model, performance in this olfactory cocktail-party problem is predicted to decline linearly as the complexity of odors increases, until there are about $27$  odorants.  Then, there will be a sharp fall-off in  probability of correct detection, approaching chance for odors with  $\sim 37$ components.  These predictions depend weakly on the number of odorants  (SI Fig.~S5 \cite{suppMat}) and strongly on the number of receptors.  Thus, if we consider a system with  $\sim300$ receptors, like human, and assuming similar decision noise, our model predicts performance  will be much worse, declining linearly until about $5-8$ components ($s = 0.05-0.10$),  then falling sharply to chance  at about $14$ components (SI Fig.~S6 \cite{suppMat}).

We compared these predictions with linear classifiers trained on  receptor responses to report whether a target odorant is present.  We  calculated  responses of $N_{\rm R} = 1000$ receptors in the CB model to random $K$-component mixtures  drawn from  $N_{\rm L}=10000$ odorants, and trained the classifier on 1000 random mixtures, half containing the target. After training, we estimated classifier performance over 1000 test mixtures, half containing the target, and averaged over 10 random sensitivity matrices, each with different sets of training/test data. This classifier's performance also declined linearly  with odor complexity (Fig.~\ref{fig:Test}d, solid and dashed black lines) but dropped earlier to chance.  The RMSE of the linear classifier and the behavioral experiment  was 0.1480, higher than our  model (0.0075).

\subsection{Neural network implementation}

\begin{figure*}
\begin{center} 
\includegraphics[width=\linewidth]{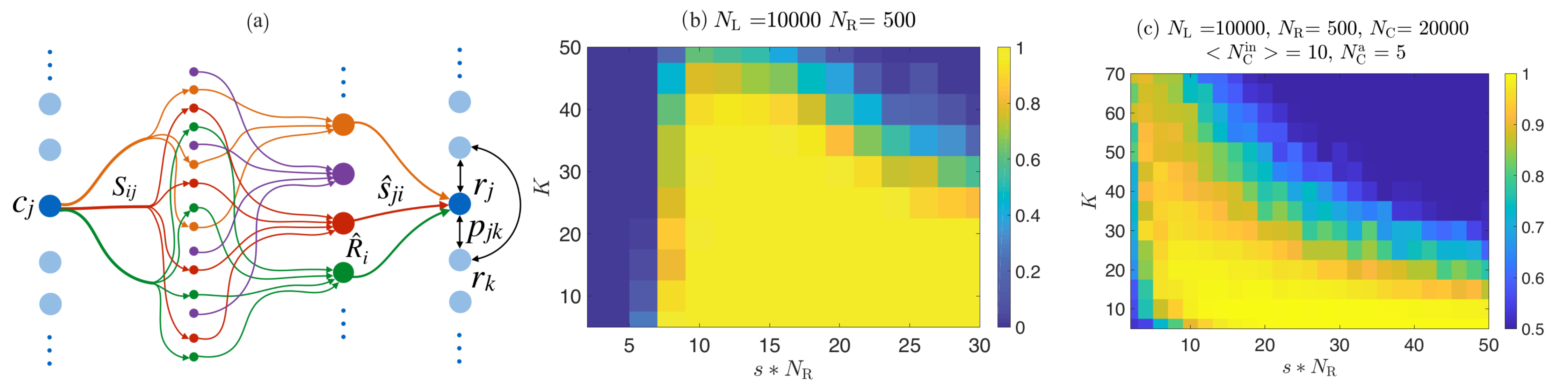} 
\caption{{\bf Network decoder:}
(a) Odorant $c_j$ binds many receptor types (colors).  Responses are reliably estimated by averaging multiple receptors of the same type in glomeruli of a second layer where axons of each type converge. Second layer inhibitory interneurons shut down  outputs of weakly activated glomeruli.  Above-threshold  responses are relayed to a readout layer whose units also receive recurrent inhibition from other readout units. Connections for one odorant and readout unit are shown. (b) Probability of correct decoding ($P({\bf r} = {\boldsymbol{c}})$)  as a function of odor complexity $K$ and $s*N_{\rm R}$ for $N_{\rm R}$ receptors and $N_L$ odorants, with $s$ =  Probability of odorant-receptor binding. 
$P({\bf r} = {\boldsymbol{c}})$  calculated numerically over 100 trials,  with  random odor mixtures and sensitivity matrices (SI: Numerical simulations \cite{suppMat}).  Correct decoding:  Euclidean distance between  odor  $\boldsymbol{c}$ and  decoded vector ${\bf{r}}$ is $<0.01$. (c) Probability of correct classification of presence/absence of a single odorant from the distributed population response in a network with random projection to a third layer with $N_C = 20000$ units, each of which has an average of 10 inputs ($\langle N_C^{\rm in} \rangle = 10)$, at least half of which must be active ($f=0.5$) to generate a response. This means that on average the third layer units need at least $N_C^a = f \langle N_C^{\rm in} \rangle = 5$  active inputs to produce a response. We used a standard linear classifier, trained with 1000 random odors, half of which contained the target odorant.  The classifier was tested on 1000 novel odors.  The classifier in panel (c) shows good performance up to higher odor complexities than in panel (b) because  it involves a simpler task -- i.e. identification of a single odorant, rather than simultaneous identification of all odor components.
}
\label{fig:NeuralModel}
  \end{center} 
  \end{figure*}

To implement our  algorithm in networks acting on realistic receptors binding stochastically to  molecules, we consider a first layer with receptor responses $R_i$ controlled by affinities $S_{ij}$ between receptors $i$ and odorants $j$  (concentrations = $c_j$).
To mitigate  noise we replicate receptors and aggregate  activity in a second layer.   Reliable non-responses are especially important for us, so we suppress responses by a standard  mechanism -- recurrent inhibition in this second layer  (Fig.~\ref{fig:NeuralModel}a) --  helping to drive weak, noisy responses below the threshold for activating the next layer of the network. This architecture parallels the  olfactory pathway, where each receptor type is individually expressed in thousands of Olfactory Sensory Neurons (Olfactory Receptor Neurons in insects), which are pooled in glomeruli of the olfactory bulb  (antennal lobe in insects). The activity of individual mitral cell outputs of each glomerulus (projection neurons in insects) is then suppressed by a widespread inhibitory network of granule cells by an amount that depends on the overall activity of all the receptors \cite{olsen2010divisive, roland2016massive} (SI Fig.~S7 \cite{suppMat}).

We feed second layer outputs ($\hat{R}$) forward with weights $\hat{S}_{ji}$ to $N_C$ third layer units (Fig.~\ref{fig:NeuralModel}a). The input to the $j^{\rm th}$ unit is gated to implement the elimination step: it is $F_j = \sum_i \hat{S}_{ji} \hat{R}_i$ if more than $f$  projections to the $j^{\rm th}$ unit are non-zero, and vanishes otherwise. This parallels the gating of the projection of the  second stage of the olfactory system  to  Piriform Cortex (Mushroom Body in insects), so that cortical neurons only respond  when many inputs are active together \cite{miyamichi2011cortical, davison2011neural, johnson2000new, franks2011recurrent},

The third layer forms a recurrent inhibitory network (weights $p_{jk}<0$) with dynamics implementing odor estimation  (Fig.~\ref{fig:NeuralModel}a).  The linearized dynamics of units with non-zero gated input  is described by
\begin{equation}
\frac{dr_j}{dt}=  -r_j + 
\sum_i \hat{S}_{ji} \hat{R}_i
+ \sum\limits_{k=1, k\ne j}^{N_{\rm C}} p_{jk}r_k \, ,
\label{eq:dynamic2}
\end{equation}
where $r_j$ are responses, and the first term on the right  describes activity decay  without inputs. $\hat{S}$ has been restricted to columns and rows associated to active receptors and readout units. Abstractly, the steady state response  representing the decoded odor satisfies: 
\begin{equation}
(\mathbb{I}-p) r =
\hat{S} \hat{R} 
\end{equation}
where $\mathbb{I}$ is the identity; $p$ and $\hat{S}$ are  recurrent/feedforward weight matrices; and  $\hat{R}$ and $r$ are response vectors for active receptors and readout units.   Thus, the steady state output linearly transforms the gated receptor response.

To illustrate the roles of the feedforward, recurrent and gating structures, suppose the number of readout units and odorants is equal ($N_C=N_L$),  sensing is linear with low noise ($\hat{R}_i = R_i = \sum_j S_{ij} c_j$), and that gating requires most projections to a responsive unit to be non-zero.  Then, at steady state, active units satisfy
\begin{equation}
(\mathbb{I}-p) r  = 
\hat{S} S c 
\end{equation}
where  rows of the odor vector $c$ and  columns of the sensing matrix $S$ have been restricted to present odorants.  This readout can directly represent odorants ($r = c$) if  $(\mathbb{I} - p)^{-1} \hat{S} = S^{-1}$,
an explicit decoding of the sort considered in \cite{penker2020mixture}.  Such an inversion of a rectangular matrix is generally ill-defined because $S$ and $\hat{S}$ will have rank less than the number of concentrations to estimate when there are fewer receptors than odorants.  But for sufficiently sparse odors and sensing matrices, we showed that the elimination step removes enough candidates from consideration to give a well defined problem -- in our network $S$ and $\hat{S}$ in the active unit dynamics will have sufficient rank to permit the inversion.

Recurrent inhibition allows local solutions to this inversion problem when it is well-defined if we choose
\begin{equation}
\sum\limits_{i=1}^{N_{\rm R}} \hat{S}_{ji}S_{ij} = 1 ~~ \mathrm{and} ~~  p_{jk} + \sum\limits_{i=1}^{N_{\rm R}} \hat{S}_{ji} S_{ik}  = 0  \ (k \neq j ) \, ,
\label{eq:weightconstraints}
\end{equation}
so that $(\mathbb{I}-p) = \hat{S} S $. 
These  criteria relate feed-forward weights to the sensing matrix recalling  \cite{zhang2016robust,penker2020mixture}, and balance the network unit by unit, compensating feed-forward excitation  by recurrent inhibition.   Since these constraints relate individual readouts (rows of $\hat{S}$) and  odorants (columns of $S$), solutions for different pairs can be spliced  to construct  feedforward and recurrent weight matrices. 
The balance criterion recalls olfactory cortex where distance-independent projections from pyramidal cells to local inhibitory interneurons produce long-range inhibition 
\cite{johnson2000new, franks2011recurrent, bathellier2009properties}, and the  Mushroom Body in insects where a giant interneuron provides  recurrent inhibition.
The $N_{\rm L} +  N_{\rm L}(N_{\rm L} -1)$ equations in (\ref{eq:weightconstraints}) are solvable because we have more parameters than  constraints: there are  $\sim s*N_{\rm R}N_{\rm L}$ feed-forward  and $N_{\rm L}(N_{\rm L}-1)$ recurrent parameters in $\hat{S}$ and $p$.  If  responses $R_i$ are nonlinear, there will still be enough parameters for decoding, but nonlinear units or multi-layer networks may be needed.  

To test the network, we selected sensitivity matrices $S$ where odorants bound randomly to a fraction $s$ of receptors, and assumed linear responses (Eq.~\ref{eq:compBind} with $d=0$), representing  statistically stable averages over many  receptors.  We then solved Eq.~\ref{eq:weightconstraints} to find feedforward and recurrent weights (SI: Numerical simulations \cite{suppMat}).  Imitating  gating of projections to  olfactory cortex \cite{miyamichi2011cortical, davison2011neural, johnson2000new, franks2011recurrent}, readout units responded if more than $95\%$ of their feed-forward inputs were active.  Fig.~\ref{fig:NeuralModel}b shows that the network performs similarly to the abstract decoders above.

The architecture above shares  features with the olfactory pathway: diffuse but sparse odorant-receptor binding; aggregation and thresholding of noisy responses in the second stage;  expansive and strongly gated projections to a recurrent network in the third stage.  But the brain does not know the number of odorants or sensing matrix, and so cannot  embody networks in which these parameters control the number of neurons or connection weights, at least without learning.  However,  a variant  algorithm works without knowing these parameters.
Suppose the feedforward  weights to the third layer are sparse, expansive, and statistically random.  Strong gating of these projections selects readout units that sample many simultaneously active inputs.  Thus, each odorant associates to a sparse readout set, whose activity will be  shaped by the dynamics  (Eq.~\ref{eq:dynamic2}) to form a distributed odor representation.

This randomly structured network will produce faithful, sparse representations of  odor mixtures in the same parameter regime as the elimination-estimation algorithm. To see this, consider a readout population  whose activity reflects the presence of odorant $j$. Because of the strong gating, each unit in this population must have a large fraction of its inputs drawn from receptors that respond to $j$.  
If   $j$ is absent in a K-component mixture, the probability that a receptor binding $j$ remains inactive is
$\sim e^{-sK}$
(SI: Eq.~S15 \cite{suppMat}).  So, of the roughly $sN_{\rm R}$ receptor types that bind to $j$, nearly $sN_{\rm R}e^{-sK}$ will be inactive. Taking typical numbers $\{K, N_R,s\} = \{10, 500, 0.05\}$,  $\sim25$ receptor types (and the corresponding second stage outputs) will respond to a given odorant, and about 60\% of these ($\sim15$) will be silent if the odorant is absent. The remaining 40\% ($\sim10$) will  respond because of other odorants in the mixture, along with additional receptor types  responsive to those odorants.  Projecting this activity randomly to the third layer,  units in the population representing $j$ that sample from  silent receptors will be inactive unless sufficiently many new inputs are active because of  other odorants in the mixture. This is unlikely because, as discussed above, the strong gating implies that units responding to  $j$ will have most of their inputs drawn from receptors that do bind to $j$.   According to our estimate about 60\% of these will be silent if $j$ is absent, despite the presence of other odorants in a mixture. Thus the readout unit will not respond. The silenced readout units thus represent  {\it absence} of odorant $j$, which can be explicitly reported by a downstream classifier trained on the sparse third stage activity.

To test this reasoning we constructed a network as described above with statistically random projections to the third layer, and trained a classifier to identify presence of a single odorant based on the third layer population response (Fig.~\ref{fig:NeuralModel}c).  We found that the classifier showed excellent performance following sparse sensing of odor mixtures with a few tens of components.  The classifier in Fig.~\ref{fig:NeuralModel}c performs well for odors with higher complexity than in Fig.~\ref{fig:NeuralModel}b, because it is performing a simpler task -- i.e. detecting a single odorant.  Similar classifiers can be built for each odorant of interest, thus forming a classifier layer that explicitly identifies the  components of a mixture.  A comprehensive future study could also explore, e.g., odor landscapes with different numbers of components, concentration ranges, and statistics; model cortices of different sizes; different statistics and gating in the projections from the second to the third stage; and different kinds of classifiers.

Similar to this decoder, projections from the olfactory bulb to the cortex seem to be statistically random \cite{caron2013random} rather than structured, and give rise to a sparse, distributed representation of odors in cortex \cite{stettler2009representations}, as opposed to a literal decoding of  odorant concentrations.  Some authors have proposed that the random projections to cortex are a mechanism for creating sparse, high dimensional representations suitable for downstream linear classification \cite{caron2013random,babadi2014sparseness,assisi2020optimality}, or are evidence for compressive sensing in olfaction \cite{stevens2015fly,zhang2016robust}.  Others have suggested that compressive sensing occurs at the receptors \cite{qin2019optimal,krishnamurthy2017disorder}, and that the random projections reformat the compressed data for downstream decoding \cite{krishnamurthy2017disorder}.  We propose a complementary view:  random projections combine with strong gating to leverage  information in silent receptors, enabling network decoding of responses from a small number of receptors.

\section{Discussion}
Our central idea is that receptors which do not respond to an odor convey far more information than receptors that do.  This is because the olfactory code is combinatorial -- each receptor binds to many different odorants and each odorant binds to many receptors.  Hence, an inactive receptor indicates that all the odorants that could have bound to it must be absent.   Natural odors are mixtures of perhaps $10$-$40$ components drawn from more than $10^4$ volatile molecules in nature  \cite{dunkel2008superscent,touhara2009sensing,yu2015drawing}.  If most of these molecules bind to a fraction of the receptors that is neither too small nor too large, odorants that are absent from a mixture can be accurately eliminated from consideration  by a system with just a few dozen to a few hundred  receptor types.   The response of the active receptors can then be used to  decode the concentrations of   molecules that are present.   Our results show that odors of natural complexity can be  encoded in, and decoded from, signals of a relatively small number of receptor types  each binding to 5-15\% of odorants. 
Perhaps this observation has a bearing on why all animals express $\sim 300$ receptor types, give or take an O(1) factor, although receptor diversity does increase in larger animals \cite{Tesileanu255547} along with the number of neurons in each olfactory structure, the latter scaling with body size \cite{srinivasan2019scaling}. Even at the extremes, the fruitfly 
 and the billion-fold heavier African elephant have $57 \sim 300/6$ \cite{vosshall2000olfactory} and $1948 \sim 300 \times 6$ \cite{niimura2014extreme} receptor types respectively.

Our network model, structured similarly to  early olfactory pathways, behaves like the abstract algorithms we proposed.  Our algorithm and   network both show  best performance if each of a few dozen to a few hundred receptor types binds to $\sim 5-15\%$ of odorants.
This is consistent with observations from {\it Drosophila} to human \cite{munch2016door, mainland2015human}.
Next, our network, like the olfactory system \cite{olsen2010divisive, roland2016massive},  pools  receptors of each type into ``glomeruli'', and  uses lateral inhibition to suppress noise activity. This achieves both reliable responses and non-responses, as required in the elimination step of our algorithm. Our network's third stage has strongly gated units  pooling  many glomeruli, most of which must be active to produce responses. The readout units also have large-scale, recurrent, balanced inhibition,  like the olfactory cortex \cite{miyamichi2011cortical, davison2011neural, johnson2000new, franks2011recurrent, bathellier2009properties}. Previous work has highlighted that  such architectures could enable robust feed-forward odor classification or reconstruction of compressed odor codes \cite{caron2013random,babadi2014sparseness,stevens2015fly,zhang2016robust,assisi2020optimality}, and supports both similarity search \cite{dasgupta2017neural} and novelty detection \cite{dasgupta2018neural}. We suggest another role for the  circuit: to use information in silence to reconstruct odor composition. We also argue that in the absence of information about the dimension of odor space and the  sensing matrix, the essential features of our algorithm could be implemented by having random, sparse projections between the second layer and a strongly gated third layer, as seen in the brain \cite{caron2013random,stettler2009representations,miyamichi2011cortical, davison2011neural, johnson2000new, franks2011recurrent}.

The latter perspective  involves a subtle point of what it means to ``decode” an odor.  Often we think of decoding as restoration of the ``original'' signal.  We are using the odorant concentrations as the “original” representation, but could instead think of clouds in molecular shape space, or a points in a space of chemical or biophysical descriptors.  Thus, the perspective that odor decoding involves direct recovery of the concentration vector is likely simplistic.  Similarly, in vision if a region of the brain decodes the presence of a cat in an image, it does not recover the actual cat, but rather a representation of ``catness'' that is easy to read.  In other words, ``decoding” essentially involves rewriting information into an easy-to-read format that can be used to generate actions.  Thus,  the random projections to cortex along with the strong gating (elimination) and recurrent activity (estimation) should be regarded as a population decoding of combinatorial odor information in receptor activity.

We discussed the steady state dynamics of our network decoder, but animal sensing is a highly dynamic affair involving sniffing, active sensing, and transient encounters with odor plumes. In this context, classic work has discussed the role of both oscillatory and transient dynamics in the brain in odor coding and decoding \cite{kay2006information,laurent1996dynamical,laurent2002olfactory,laurent2001odor,mazor2005transient,rabinovich2000dynamical,turner2008olfactory,bazhenov2009olfactory,assisi2020optimality}. It will be interesting for the future to study how these dynamics interact with the combinatorics of silence that we have discussed, along with the extensive learning and plasticity that occur in the olfactory system.

Future work could also include and study the role of inhibitory and suppressive interactions that have been noted in the nose, where an odorant which does not activate a particular receptor instead suppresses responses of that receptor to other odorants \cite{reddy2018antagonism,pfister2019odorant,zak2020antagonistic}.  Odor-invoked inhibitory responses could be formally included in our competitive binding model by including negative entries in the sensing affinity matrix.  In this case, the weak response of an ORN may mean that odorants that activate the receptor are absent, or that both excitatory and inhibitory odorants are present at high concentrations. There are potential strategies for including such suppression in our algorithm if odor-receptor interactions are sparse as experiments suggest \cite{saito2009odor, mainland2015human}. Specifically, in the algorithms that assumed knowledge of the receptor response model, we would ignore odorants and receptors  with strong suppressive interactions  in the elimination step, and  include them in the estimation step.   If the network decoder does not assume knowledge of the odorant-receptor affinities and instead uses random projections, the readout classifier would have to learn these modifications.  We leave detailed analysis for the future.

Our model suggests that estimation and discrimination of complex odors should improve with the size of the  receptor repertoire. For example, while humans, dogs and mice might perform similarly at low odor complexity,  the latter animals should be better than humans at discriminating more complex odors as they have 2.5 times more  receptor types. The quantitative predictions for our model can be determined by studying odor discrimination thresholds as a function of odor complexity for receptor repertoires of different sizes.

Our model also suggests that the information needed for discriminating  odor composition may be present in  combinatorial receptor representations, contrary to our usual experience of olfaction as a synthetic sense.  In fact  experiments do show that complex odors  differing by just a few components can be discriminated in some circumstances \cite{jinks1999limit, bushdid2014humans}.  If the principles underpinning our  algorithm are reflected in the brain, as suggested by the analogy with our network model,  odors that bind to inactive receptor types should be largely eliminated.  We could test this by blocking receptor types pharmacologically, or via optogenetic suppression, and expect that animals will  behave as if odorants binding to suppressed receptors are absent, even if other receptors do bind them.  Finally, in our model, odors can be decoded well (yellow regions in Figs.~\ref{fig:binaryScheme},\ref{fig:continuousScheme},\ref{fig:NeuralModel})  if they have fewer than $K_{{\rm max}}$ components, where $K_{{\rm max}}$ is determined by the number of receptor types ($N_{\rm R}$) and the fraction of them that bind to typical odorants  ($s$).  We can test this  by measuring $N_{\rm R}$ and $s$ for different species and  characterizing discrimination performance between odors of complexity bigger and smaller than $K_{{\rm max}}$ (see Fig.~\ref{fig:Test}).

Our algorithm differs from compressed sensing \cite{candes2006robust, candes2005decoding, donoho2006compressed, ganguli2012compressed} which uses a dense, linear sensing matrix to represent high-dimensional sparse vectors in a low dimensional signal, which is decoded through constrained minimization.
We similarly assume sparsity of the input vector, but we do not assume linear sensing, and
our sensing matrices are sparse, like known odorant-receptor interaction matrices \cite{mainland2015human,munch2016door}. 
Also, our decoding mechanism exploits sensory silence, rather than imposing  sparsity in the decoded vector \cite{rozell2008sparse}.  Our approach cannot provide the same general decoding guarantees as compressed sensing, but it  succeeds well in the relevant regime of parameters.

Finally, our algorithm  may have applications for decoding complex  odors detected by chemosensing devices like electric noses \cite{johnson2006dna, goldsmith2011biomimetic}.    In this engineered setting, the target odorants and response functions are explicitly known so that our method of ``Estimation by Elimination'' can be precisely implemented.

\noindent {\bf Acknowledgements:} VS was supported by the University of Pennsylvania Computational Neuroscience Initiative and NIH-SC2GM140945.
VB was supported by Simons Foundation MMLS grant 400425,  and NSF grants PHY-160761 and  PHY-1734030. VB thanks the Kavli IPMU for hospitality.  We are grateful to Vikas Bhandawat for useful communications.

\bibliography{references.bib}

\clearpage
\renewcommand{\thefigure}{S\arabic{figure}}
\setcounter{figure}{0}
\setcounter{section}{0}

\renewcommand{\theequation}{S\arabic{equation}}
\setcounter{equation}{0}
{\huge Supplementary Information}

\section{Analytic estimate of the probability of correct decoding}
\subsection{Identifying odorant presence}
We want the probability $P(\hat{\boldsymbol{c}} = \boldsymbol{c})$ that the decoded vector $\hat{\boldsymbol{c}}$ equals the input vector $\boldsymbol{c}$, i.e., the corresponding elements of the vectors $\hat{\boldsymbol{c}}$ and $\boldsymbol{c}$ are equal.   Assuming statistical independence of the decoding of each odorant, we can write

\begin{equation}
P(\hat{{\boldsymbol{c}}} = {\boldsymbol{c}}) = \prod_{i=1}^{N_{\rm L}} P(\hat{c}_i=c_i) = \left[P(\hat{c}_i=c_i) \right]^{N_{\rm L}} \, .
\label{eq:Probability1}
\end{equation}
The assumption of independence is an approximation that we will validate by comparing with the full numerical results.

The decoded concentration $\hat{c}_i$ could be equal to $c_i$, if either both of them equal 1 or both of them equal zero. 
Thus, the term in the square bracket in Eq.~\ref{eq:Probability1} can be written as:
\begin{align}
P(\hat{c}_i=c_i) & = P(\hat{c}_i=1| c_i =1) P(c_i = 1) \nonumber \\ &+ P(\hat{c}_i = 0 | c_i = 0) P(c_i = 0).
\label{eq:ProbabilityElements}
\end{align}
where $P(c_i =1) = K/{N_{\rm L}} = \alpha$ is the probability that an odorant is present in the mixture, and $P(c_i = 0) = (1-\alpha)$.

The decoder guarantees that if an odorant $c_i$ is present and there is a receptor $R_j$ that is sensitive to it ($S_{ji}$=1), then the receptor will respond, and the decoded vector will set the corresponding element $\hat{c}_i$  to 1. If no receptor is sensitive to this odorant (i.e, $\forall j~: ~j\in[1,N_{\rm R}], S_{ji}=0 $),  the decoded element will still be set to 1 by default.
So, $P(\hat{c}_i = 1 | c_i = 1)=1$.

To calculate $P(\hat{c}_i=0| c_i =0)$, recall that in our decoding scheme, $\hat{c}_i = 0$ if there exists at least one receptor such that $R_j = 0$ for which $S_{ji}=1$. Thus,
\begin{equation}
P(\hat{c}_i=0| c_i =0) = P(\exists \; j: R_j =0 \cap S_{ji}=1|c_i = 0)
\end{equation}
where $\cap$ is the binary AND operation.
The probability on the right is 1 minus the probability that for all receptors either $R_j=1$ or $R_j=0 \cap S_{ji}=0$. So, 
\begin{align}
& P(\hat{c}_i=0| c_i =0)  \nonumber \\ & = 1 - P(\forall \; j: R_j =1 \cup (R_j=0 \cap S_{ji}=0)|c_i = 0) \nonumber \\
& = 1 - \left[P(R_j =1 \cup (R_j=0 \cap S_{ji}=0)|c_i = 0)\right]^{N_{\rm R}},
\label{eq:receptorsAreIndependent}
\end{align}
where in the second step we have again made the assumption that the receptors are independent conditional on the response of $c_i$.
The quantity in the bracket in Eq.~\ref{eq:receptorsAreIndependent} can be written as:
\begin{align}
& P(R_j =1 \cup (R_j=0 \cap S_{ji}=0)|c_i = 0) \nonumber \\
& = P((R_j =1 \cup R_j=0) \cap (R_j =1 \cup S_{ji}=0)|c_i = 0) \nonumber \\
& = 1 \cap (R_j =1 \cup S_{ji}=0)|c_i = 0) \nonumber \\
& = P(R_j =1 \cup S_{ji}=0|c_i = 0) \nonumber \\ 
& = 1 - P(R_j =0 \cap S_{ji}=1|c_i = 0) \nonumber \\ 
& = 1 - P(R_j =0|c_i = 0)P(S_{ji}=1|c_i = 0)
\end{align}
Now, $P(S_{ji}=1|c_i = 0) = P(S_{ji}=1) = s$, where entries of the sensing matrix are chosen to be non-zero independently and with probability $s$.

To calculate $P(R_j =0|c_i = 0)$ recall that the receptors are OR gates with inputs $S_{jk}c_{k}$. 
Thus, for $R_j=0$ all terms $S_{jk}c_k$ should be zero. 
The probability that any one such term is zero is $(1 - s \alpha)$. 
Since we already have $c_i = 0$, there are $(N_{\rm L} -1)$ additional terms that need to be zero. 
Hence, 
\begin{equation}
P(R_j =0|c_i = 0) = (1 - s \alpha)^{(N_{\rm L} -1)} \, ,
\label{eq:probReceptorInactiveExact}
\end{equation}
and
\begin{equation}
P(\hat{c}_i=0| c_i =0) =  \left(1 - \left[1-s(1-s\alpha)^{(N_{\rm L}-1)}\right]^{N_{\rm R}}\right) 
\label{eq:zeroProbability}
\end{equation}

Putting this all together (using Eq.~\ref{eq:zeroProbability} in Eq.~\ref{eq:ProbabilityElements}), we get:
\begin{align}
&P(\hat{{\boldsymbol{c}}} = {\boldsymbol{c}})  =  \nonumber \\ & \left[ \alpha + (1-\alpha) \left(1 - \left[1-s(1-s\alpha)^{(N_{\rm L}-1)}\right]^{N_{\rm R}}\right) \right]^{N_{\rm L}}
\label{eq:correctest}
\end{align}
Using Eq.~\ref{eq:zeroProbability}, we can also get the (approximate) probability of a false detection as $P(\hat{c}_i=1| c_i =0) = 1 - P(\hat{c}_i=0| c_i =0)$:
\begin{equation}
P(\hat{c}_i=1| c_i =0) \approx \left[1-s(1-s\alpha)^{(N_{\rm L}-1)}\right]^{N_{\rm R}}.
\label{eq:flasePositiveAppend}
\end{equation}
This expression is approximate due to our independence assumptions.

\subsubsection{Approximation:}
Since the average number of odorants present in the mixture ($K = \alpha N_{\rm L}$) is small compared to $N_{\rm L}$ and $N_{\rm L} \gg 1$, we can approximate:
\begin{equation}
(1-s\alpha)^{(N_{\rm L}-1)} = \left(1-\frac{s K}{N_{\rm L}}\right)^{(N_{\rm L}-1)} \approx e^{-s K}.
\end{equation}
Now, since the odor sensitivity ($s$) is small, so that $s e^{-s K}$ is also small, while $N_{\rm R} \gg 1$, we further approximate 
\begin{equation}
\left[1 - s e^{-s K} \right]^{N_{\rm R}} = \left[1 - \frac{s N_{\rm R} e^{-s K}}{N_{\rm R}} \right]^{N_{\rm R}} \approx e^{-s N_{\rm R} e^{-s K}}.
\end{equation}
This results in:
\begin{equation}
P(\hat{{\boldsymbol{c}}} = {\boldsymbol{c}})  = \left[ \alpha + (1-\alpha) \left(1 - e^{-s N_{\rm R} e^{-s K}}\right) \right]^{N_{\rm L}}
\label{eq:BooleanProb2}
\end{equation}
which simplifies to:
\begin{equation}
P(\hat{{\boldsymbol{c}}} = {\boldsymbol{c}})  = \left[ 1 - e^{-s N_{\rm R} e^{-s K}} + \alpha e^{-s N_{\rm R} e^{-s K}} \right]^{N_{\rm L}}
\end{equation}
This expression approximates to Eq.~\ref{eq:BooleanProb4} in the main text:
\begin{equation}
P(\hat{{\boldsymbol{c}}} = {\boldsymbol{c}})  = \left[ 1 - N_{\rm L} e^{-s N_{\rm R} e^{-s K}}  \right]
\end{equation}
Similarly, Eq.~\ref{eq:probReceptorInactiveExact} approximates to
\begin{equation}
P(R_j =0|c_i = 0) = (1 - s \alpha)^{(N_{\rm L} -1)}\approx e^{-s\alpha N_{\rm L}} = e^{-sK},
\label{eq:probReceptorInactive}
\end{equation}
and
Eq.~\ref{eq:flasePositiveAppend} approximates to
\begin{equation}
P(\hat{c}_i=1| c_i =0) \approx e^{-s N_{\rm R} e^{-s K}}.
\end{equation}

\subsection{Identifying odorant concentrations}
For the continuous decoder to give a unique solution,
the number of receptors that respond to the mixture should be larger than the number 
of odorants with non-zero concentrations ($K = \alpha N_{\rm L}$). 
This ensures that the system of equations is over-determined and 
can, in principle, be solved.

Additionally, the number of receptors that do not respond should be such that the absent odorants  can be set to zero.
Since every receptor binds to $sN_{\rm L}$ odorants on average, we need at least $1/s$ receptors to cover all the odorants.
In general, as the entries of the sensitivity matrix are statistically distributed,  the number of receptors that do not respond should be larger than ${\gamma}/{s}$ for correct odor estimation, where $\gamma$ is a small number greater than 1. 

Putting this all together, if $P(\tilde{N}_{\rm R})$ is the probability of the number of receptors with non-zero response, 
we are interested in the probability that $P(\tilde{N}_{\rm R}>  K = \alpha N_{\rm L})*P(N_{\rm R}-\tilde{N}_{\rm R}> (\gamma/s))$. The probability that a receptor responds is:
\begin{equation}
P(R>0) = \left( 1 - P(R=0) \right) = \left(1- (1 - s \alpha)^{N_{\rm L}}\right).
\end{equation}
Taking the number of receptors that respond to be a Poisson 
variable with rate $\left< \tilde{N}_{\rm R}\right> = N_{\rm R}*P(R>0)$, 
we can estimate the typical number of receptors that respond.
For biologically appropriate parameters  $\{N_{\rm L}, N_{\rm R}, K, s\} \sim \{10^4, 500, 10, 0.05\}$, 
the mean number of receptors that respond is $\left< \tilde{N}_{\rm R}\right> \sim 200$.
The standard deviation is $\sqrt{\left< \tilde{N}_{\rm R}\right>} \sim 14$.
For these values of the mean and variance, we can approximate the 
Poisson distribution with a Gaussian $P(\tilde{N}_{\rm R}) = \mathcal{N}\left(\tilde{N}_{\rm R}, \sqrt{\tilde{N}_{\rm R} } \right)$.
Thus, 
\begin{align}
& P(\hat{{\boldsymbol{c}}} = {\boldsymbol{c}})  \sim P(\tilde{N}_{\rm R} > \alpha N_{\rm L})*P(N_{\rm R}-\tilde{N}_{\rm R}> (\gamma/s))
= \nonumber \\ & \left[ 1 - \Phi \left( \frac{\alpha N_{\rm L} - \tilde{N}_{\rm R}}{\sqrt{\tilde{N}}_{\rm R}}\right)\right] 
\Phi \left( \frac{N_{\rm R} - \tilde{N}_{\rm R} - \frac{\gamma}{s}}{\sqrt{\tilde{N}}_{\rm R}}\right)
\label{eq:theoreticalLimit}
\end{align}
where $\Phi$ is the cumulative distribution function of the standard normal distribution.

\subsection{Numerical Simulations}
\subsubsection{Identifying odorant presence} For the binary case, the elements of the odor vector were chosen to be non-zero with probability $P(c_i>0) = K/N_{\rm L}$. The entries of the sensitivity matrix $S_{ij}$ were chosen to be non-zero with a probability $s$, ($P(S_{ij}) = s$). The receptor response was calculated using the binary `OR' function. The decoded concentration $\hat{c}$ was estimated using the two steps described in the main paper. First, the decoded concentration of any odorant to which an inactive receptor is sensitive, was set to zero. All remaining concentrations were set to 1.

 \begin{figure}
\begin{center} 
\includegraphics[scale=0.3,trim=0cm 0.0cm 0.0cm 0.0cm]{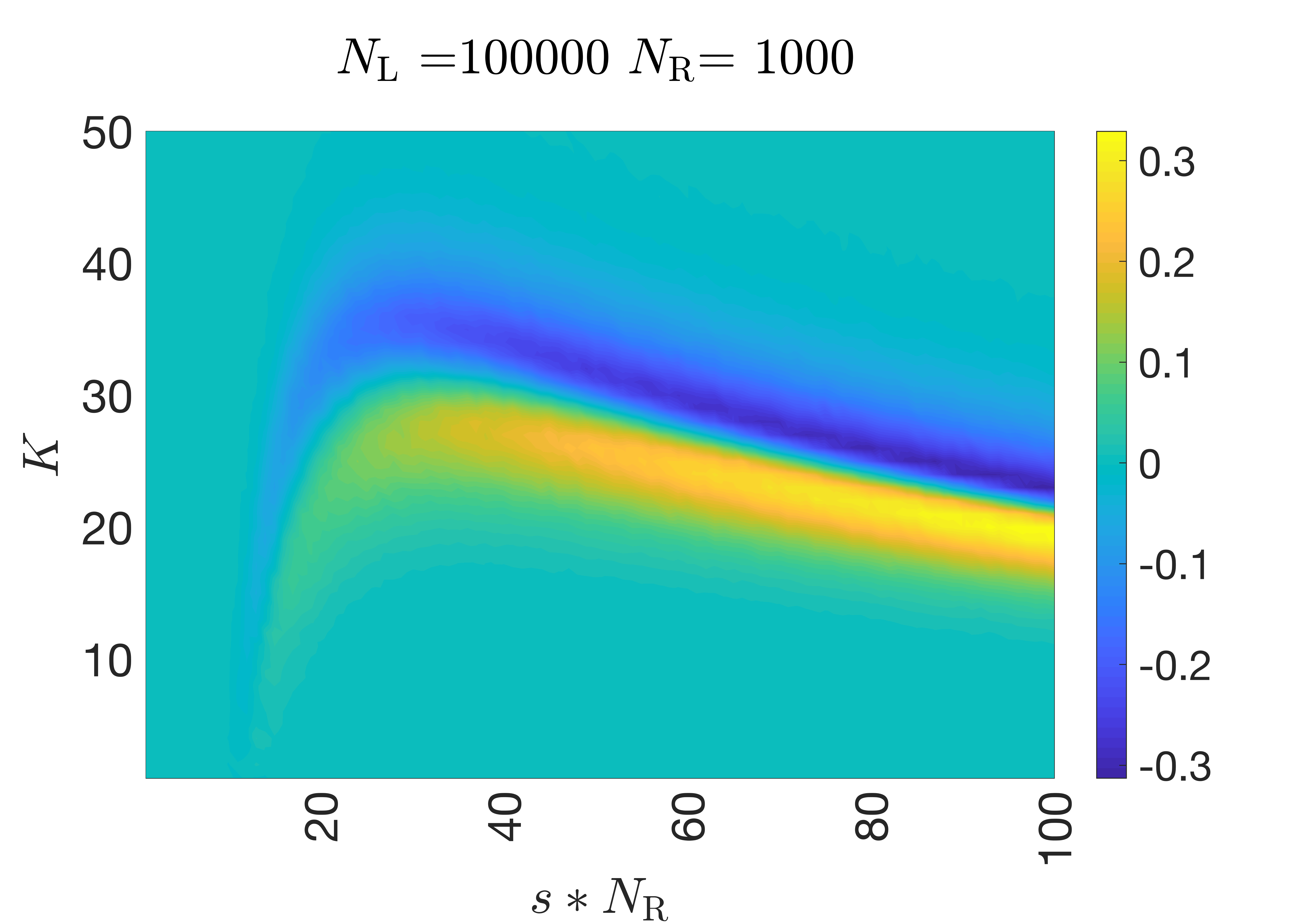} 
    \caption{Plot of the difference between $P(\hat{{\boldsymbol{c}}} = {\boldsymbol{c}})$ as given by Eq.~\ref{eq:correctest} and as estimated numerically for the binary decoder. For most parameters the analytical results match the simulations.
\label{fig:binaryNumericsVsAnalytics}}
  \end{center} 
  \end{figure}

\begin{figure*} 
\begin{center} 
\includegraphics[width=\linewidth]{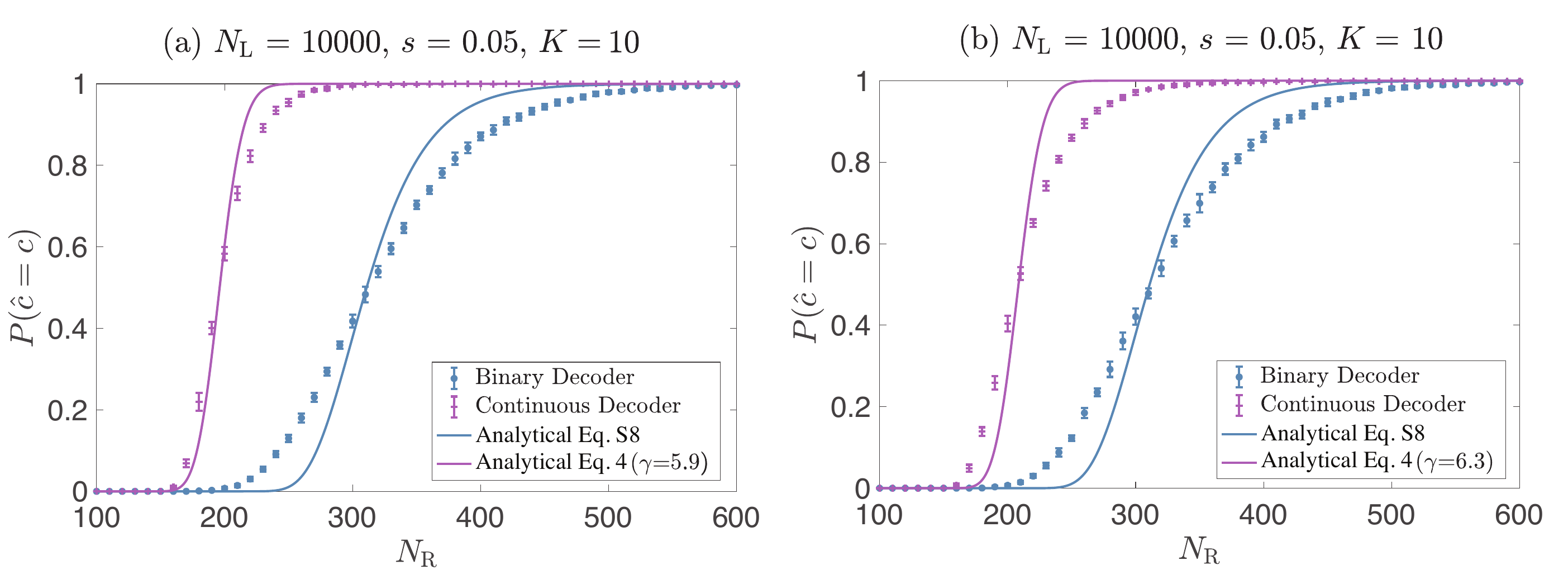} 
    \caption{{\bf $P(\boldsymbol{\hat{c}}=\boldsymbol{c})$ as a function of $N_{\rm R}$ for the continuous decoder and alternative choices of the sensitivity matrix}. Results for binary encoding are the same as in Fig.~\ref{fig:binaryScheme}c and are plotted here for comparison. (a) Uniform distribution: Similar to Fig.~\ref{fig:continuousScheme}c  except that the non-zero elements of the sensitivity matrix were chosen uniformly at random between [0,1]. (b) Log-normal distribution: Similar to Fig.~\ref{fig:continuousScheme}c except that the non-zero elements of the sensitivity matrix were chosen at random from a log normal distribution with the corresponding normal distribution having mean zero and standard deviation 1.
\label{fig:comparisonWithOtherDistributions}}
  \end{center} 
  \end{figure*}

  \begin{figure*} 
\begin{center} 
\includegraphics[width=\linewidth]{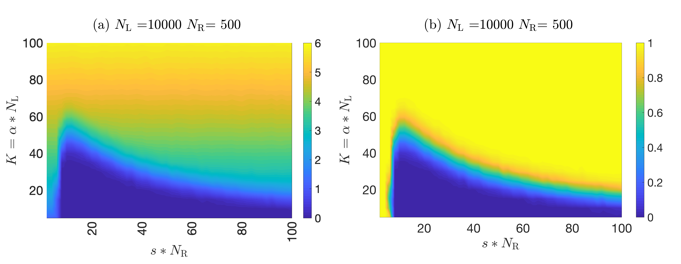} 
    \caption{{\bf Other measures of estimation error}: (a) Total error: $L_2$ norm (or the root mean square) of the difference between actual and estimated concentrations for the continuous decoder with compressive binding encoding model. (b) $L_2$ norm of the difference between actual and estimated concentrations divided by the $L_2$ norm of the actual concentration.
\label{fig:errorMeasures}}
  \end{center} 
  \end{figure*}

\begin{figure}
\begin{center} 
\includegraphics[width=0.8\linewidth]{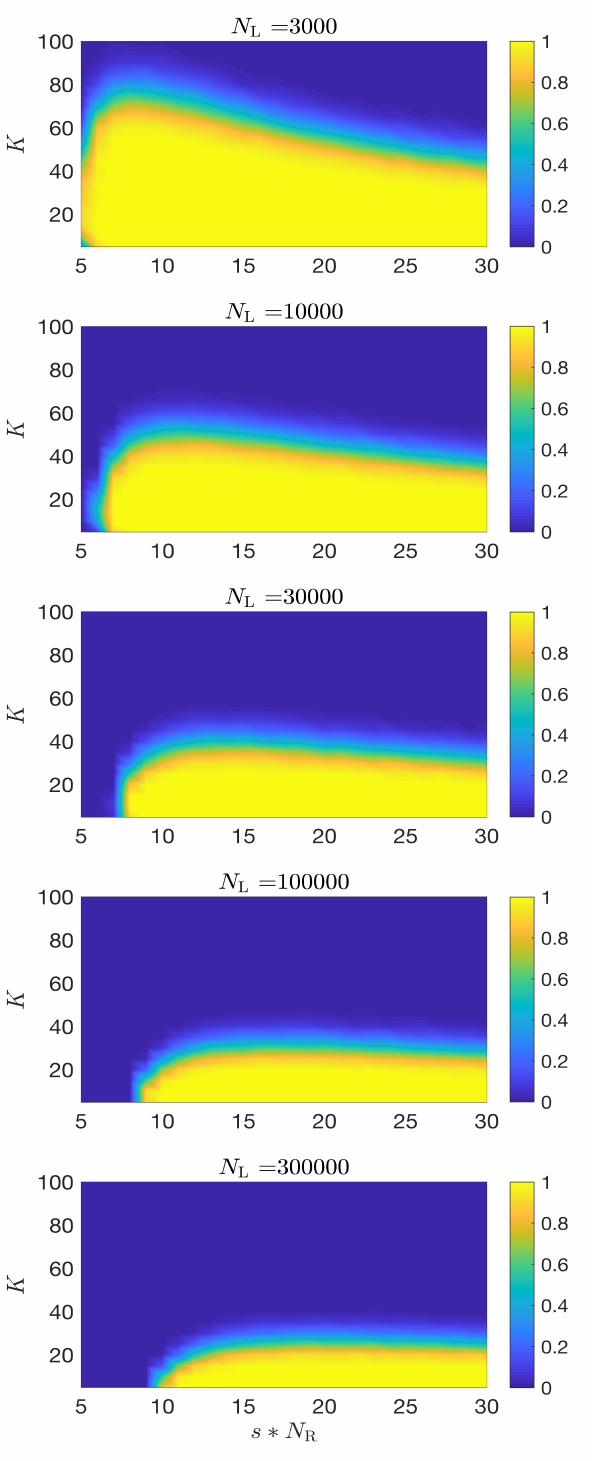} 
    \caption{{\bf Dependence of $P(\boldsymbol{\hat{c}} = \boldsymbol{c})$ on $N_{\rm L}$}: $P(\boldsymbol{\hat{c}}=\boldsymbol{c})$  plotted as a function of odor complexity $K$ and $sN_{\rm R}$ at a fixed value of number of receptors $N_{\rm R} = 500$. Each panel gives $P(\boldsymbol{\hat{c}}=\boldsymbol{c})$ for a different value of the total number of possible odorants $N_{\rm L}$.   The minimum value of $s N_R$ for successful decoding and the optimal value where the most complex odors can be decoded are both relatively independent of  $N_{\rm L}$.
\label{fig:variationWithNL}}
  \end{center} 
  \end{figure}

\subsubsection{Identifying odorant concentrations} For the continuous case, the elements of the odor vector were chosen to be non-zero with probability $P(c_i>0) = K/N_{\rm L}$, and the elements of the sensitivity matrix were chosen to be non-zero with probability $s$, ($P(S_{ij}>0) = s$). The values of the non-zero elements in the odor vector were chosen from a uniform distribution on the interval $[0, 1)$, and for the sensitivity matrix from a log-uniform distribution between $10^{-1}$ and $10^1$. 
The activity of each receptor was determined using Eq.~\ref{eq:compBind} of the main text (d = 1). 

The concentration of any odorant to which an inactive receptor is sensitive was set to zero. 
After this elimination, let $\tilde{\boldsymbol{R}}$ be the vector representing the response of the set of active receptors, 
$\tilde{\boldsymbol{c}} $ be the vector representing the concentration of the odorants that have not been set to zero, and
$\tilde{S} = \{\tilde S_{ij} \}$ be the $\tilde{N}_{\rm R} \times \tilde{N}_{\rm L}$ sensitivity submatrix over active receptors $\tilde{\boldsymbol{R}}$ and the remaining odorants $\tilde{\boldsymbol{c}}$.
Then, if $\tilde{N}_{\rm R} < \tilde{N}_{\rm L}$ (non-invertible case), all decoded concentrations were set to zero. Otherwise, the decoded concentrations were given by the vector
that minimized the $L_2$ distance $||\tilde{\boldsymbol{R}} - \tilde{S}\cdot\tilde{\boldsymbol{c}}||_2$. 
The Levenberg-Marquardt solver with geodesic acceleration from
the \textit{GNU GSL} library was used to find the minimum.

Multiple trials were run for each choice of parameters. 
At the end of each trial, the $L_2$ norm of the difference between actual and decoded concentration vectors was reported. 
The trial was considered a success if this norm was less than a threshold of $0.01$.

Simulations were performed in C++.
The sensitivity matrix $S$ and the odorant concentrations were generated from streams of (pseudo)random numbers drawn by the \textit{Xoroshiro128+} random number generator. 
Each stream is seeded with a $2^{64}$ forward jump from the seed of the previous stream. 
The first stream is seeded from the output of the \textit{SplitMix64} generator initialized by current system time. 
Random number production as well as vector operation code were optimized using \textit{Intel}'s SIMD instruction set.

\subsubsection{Neural network} To simulate the neural network, we generated random sparse odor vectors and sensitivity matrices. The elements of the odor vector were chosen to be non-zero with probability $P(c_i>0) = K/N_{\rm L}$, and the elements of the sensitivity matrix were chosen to be non-zero with probability $P(S_{ij})>0 = s$. The value of the non-zero elements were chosen from a uniform distribution on the interval $[0, 1)$. The receptor response was calculated using a linear response model (d = 0 in Eq.~\ref{eq:compBind}).
To get the feed forward connections $\hat{S}_{ji}$, we first made a matrix $\bar{\mathcal{S}}$ defined as: $\bar{\mathcal{S}}_{ji} = 1/(S_{ji})$ if $S_{ji}$ is non-zero, and $\bar{\mathcal{S}}_{ji} = 0$ otherwise. The matrix $\hat{S}$ was then chosen as: $\hat{S}_{ji} = \bar{\mathcal{S}}_{ji}/ \left(\sum \limits_{i}\bar{\mathcal{S}}_{ji}S_{ij}\right)$. The elements of the recurrent connectivity matrix were obtained as $p_{jk} = - \sum \limits_{i} \hat{S}_{ji} S_{ik}$. 

If more than 5\% of the receptors connected to a readout unit $r_{j}$ were inactive, the decoded concentration $r_j$ was set to zero. 
The feed-forward input to the remaining readout units were calculated as $r^{\rm init}_j = \hat{S}_{ji}\hat{R_i}$.
The remaining concentrations were computed as $(1-\tilde{p})^{-1} \mathbf{r}^{\rm init}$, where $\mathbf{r}^{\rm init}$ is the vector representing the total feed-forward input to neurons that have more than 95\% of their receptors active, and $\tilde{p}$ represents the sub-matrix of  connection weights between these neurons.

\subsubsection{Network Classifier}

To simulate the network classifier, we generated random sparse odor vectors and sensitivity matrices similarly to the neural network described above. The receptor/glomerular responses were calculated using the competitive binding model. The receptor/glomerular response was projected randomly to a third layer (cortical layer with $N_{\rm C}$ neurons): the connectivity matrix was generated so that each second layer unit had a small nonzero probability of connection to each third layer unit such that third layer units received an average of $\langle N_{\rm C}^{\rm in} \rangle$ inputs. The connection strengths were sampled from a uniform distribution. A unit in the third layer produced a response if more than a fraction $f$ of its inputs was active. If active, the response was calculated using a saturating activation function of the form given in (Eq. 3).
For the simulations in the main text, we chose $\langle N_{\rm C}^{\rm in} \rangle = 10$ and $f = 0.5$. Thus, for a neuron in the third layer to respond, on average at least $N_{\rm C}^{\rm a} = f*\langle N_{\rm C}^{\rm in}\rangle = 5$ of the inputs needed to be active. We built a linear classifier based on the response of the neurons in the third layer, using the Matlab  function {\em fitclinear}.

To train the classifier, we generated 1000 random sparse odor vectors, half of them containing a specific odorant and the other half not. We calculated the response of the neurons in the third layer for these random odors. We then tested the performance of the classifier on a test set of 1000 different random sparse odors, half of which contained the specific odorant. We calculated fraction of times the classifier correctly identified the odorant to be present or absent. 

We repeated this process 10 times, each with a new sample of the odor-receptor sensitivity matrix and receptor-cortex projection matrix. We estimated classifier performance as the fraction of correct identification averaged over these 10 simulations. For the plot presented in Figure 4c of the main text, we repeated the above process  for each combination of the parameters $K$ and $s*N_{\rm R}$ with new samples of sensitivity and projection matrices.

\begin{figure*}
\begin{center} 
\includegraphics[width=\linewidth]{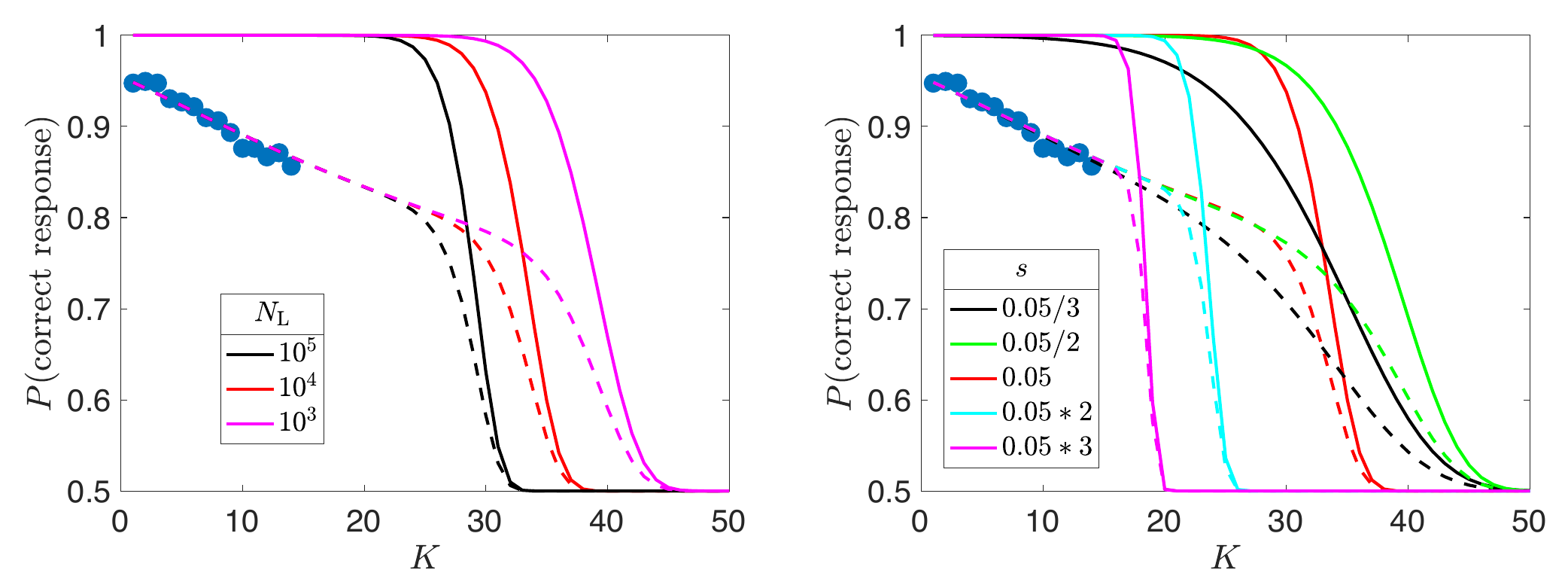} 
    \caption{{\bf Effect of the number of odorants $N_{\rm L}$ and receptor sensitivity $s$ on performance in an olfactory cocktail party problem}:   Probability of correct detection of presence or absence of an odorant in a $K$-component mixture. Compare to Figure~\ref{fig:Test}d. (a) $N_{\rm R} = 10^3, s = 0.05$. (b) $N_{\rm R} = 10^3, N_{\rm L} = 10^4$.
    }
  \end{center} 
  \label{sifig:numodorants}
  \end{figure*}

\begin{figure*}
\begin{center} 
\includegraphics[width=0.6\linewidth]{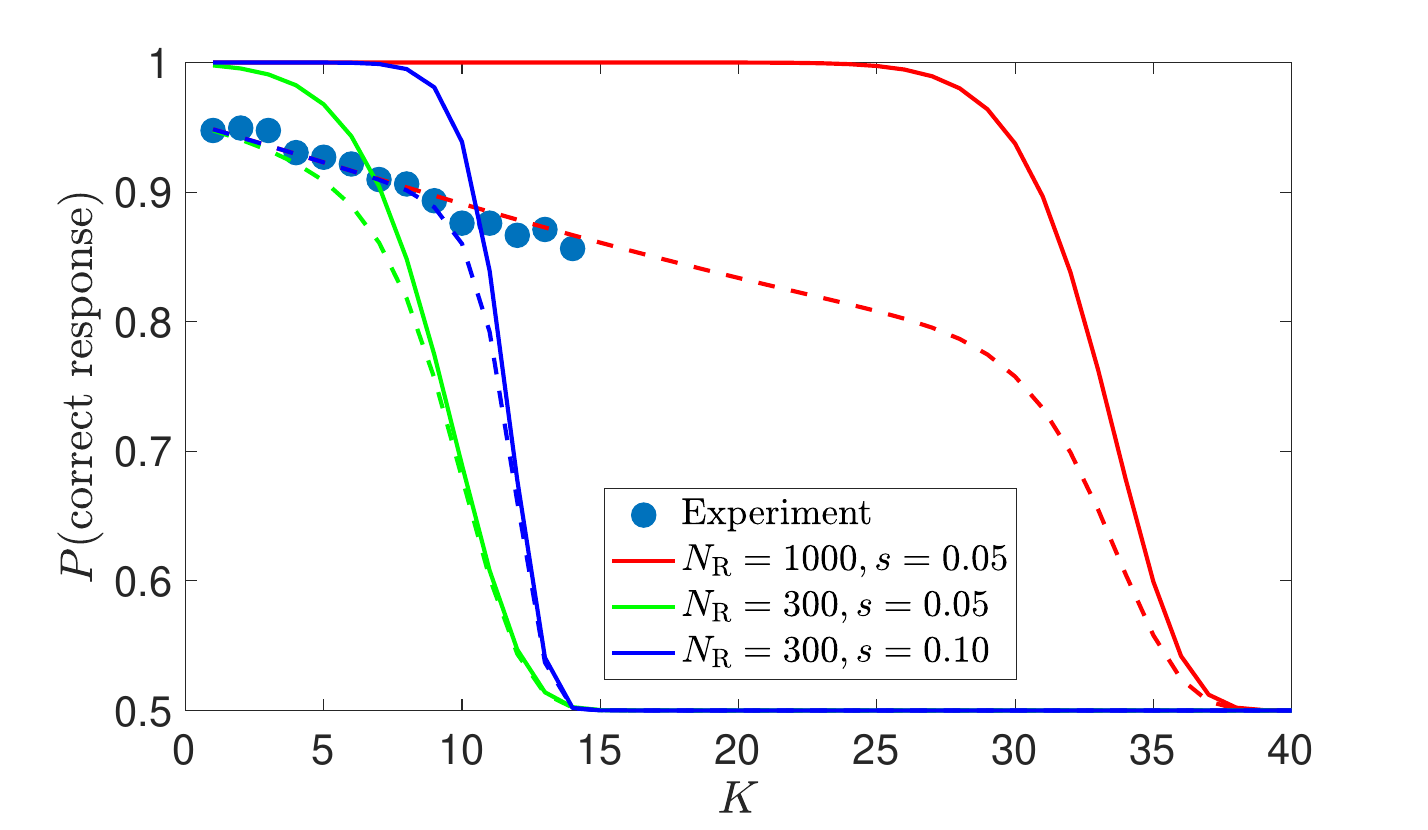} 
    \caption{ {\bf Effect of number of receptors on performance in an olfactory cocktail party problem}: Probability of correct detection of presence or absence of an odorant in a $K$-component mixture. Blue markers =  fraction of correct responses  (true positive + correct rejection) by mice.  Continuous lines = prediction in the absence of noise for the binary decoder. Dashed lines = prediction including noise determined from the data in \cite{rokni2014olfactory} (Fig. 3c main text). Performance with $N_{\rm R} \sim 300 $ receptors, like human, is predicted to be lower than with  $N_{\rm R} \sim 1000$ receptors, like mouse, falling to chance level at $K\sim 14$. We have shown the prediction with $300$ receptors at two values of the receptor-odor sensitivity ($s = 0.05$ (green lines) and $s = 0.10$ (blue lines)).
    } 
  \end{center} 
  \label{sifig:human-mouse}
  \end{figure*}

\begin{figure*}
\begin{center} 
\includegraphics[width=0.5\linewidth]{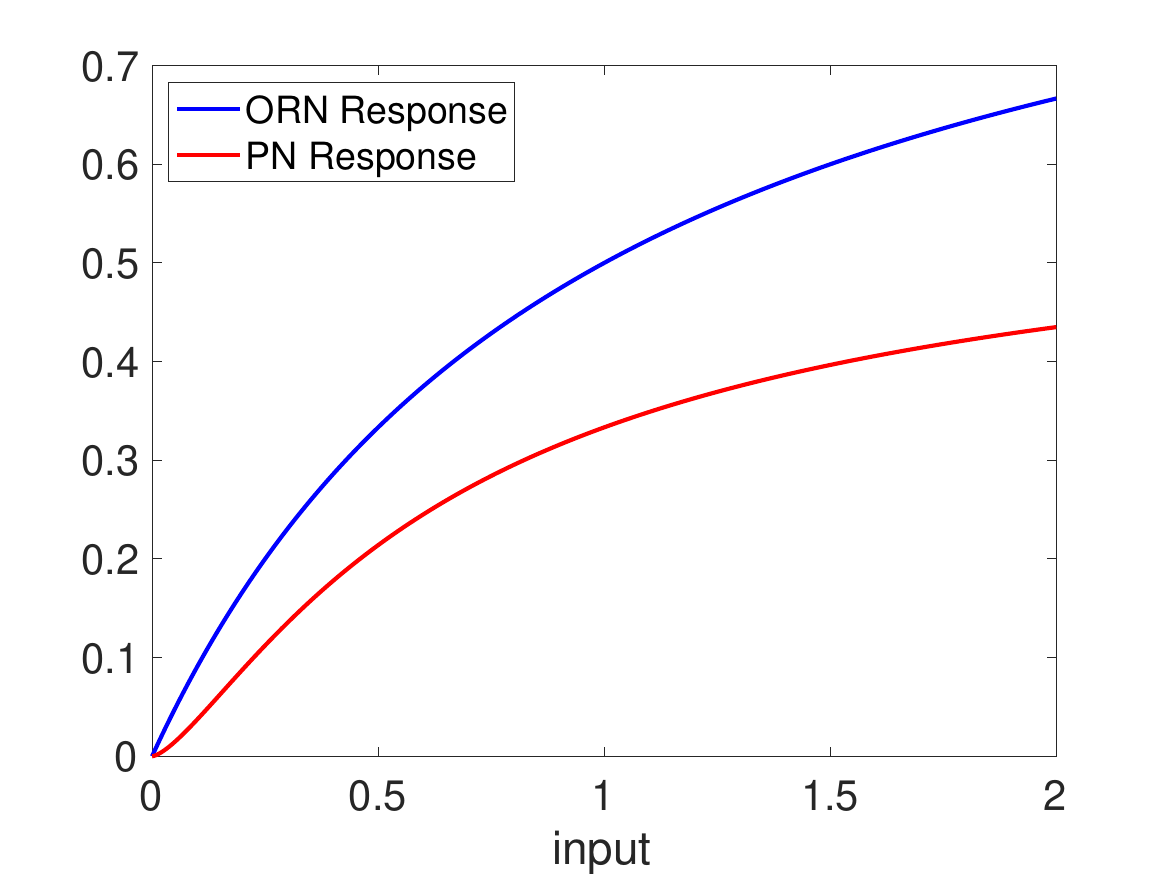} 
    \caption{{\bf Effect of divisive normalization on output neurons of Antennal Lobe (analog of Olfactory Bulb in insects)}: The response of the output Projection Neurons (PN) of the Antennal Lobe undergoes a divisive normalization driven by lateral inhibition \cite{olsen2010divisive}.  The PN output is related to the Olfactory Receptor Neuron (ORN) activity   by $PN = R_{\rm max} \frac{(ORN)^n}{(ORN)^n + \sigma^n + s^n}$, where $\sigma$ parametrizes spontaneous activity, $s$ increases in proportion to the overall activity of ORNs, and the coefficient $n = 1.5$. $R_{\rm max}$ is a scale determined by the amount of ORN pooling in the glomeruli of the Antennal Lobe.   Here we set $R_{\rm max} = 1$ because the overall scale is not relevant to our considerations; rather we are interested in the relative suppression due to overall activity of receptors.
    Since $n>1$, the response of each projection neurons is suppressed by the overall activity.  While all PNs are suppressed, weakly active glomeruli will be driven closer to, or below, activation thresholds for the next stage of processing.   In this figure, the ORN response is given by the competitive binding model, and we chose $\sigma = 1$, $s=1$ in arbitrary units to illustrate the suppression. Note that  in \cite{olsen2010divisive}  the measured values were $\sigma = 12$ spikes/s, $s$ in the range 0-50 spikes/s, and $R_{{\rm max}} \sim 150$ spikes/s (Private Communication with authors of \cite{olsen2010divisive} correcting a typographical error in the printed paper).   The authors of \cite{olsen2010divisive} used mixtures of only two odors at low concentrations in order to avoid widespread activation of ORNs, and hence the measured values of $s$ in their experiments are likely low, suggesting  stronger suppression in natural conditions (Private Communiation with authors of \cite{olsen2010divisive}).
    } 
  \end{center} 
  \label{sifg:divisive}
  \end{figure*}

\end{document}